\documentclass[usegraphicx,usenatbib]{mn2e}
\usepackage{amssymb}
\usepackage[dvips]{graphicx}
\usepackage{longtable}
\usepackage{epstopdf}
\def\Teff{$T_{\rm eff}$}
\def\logg{$\log\,g$}

\def\Vt{V${\rm t}$}
\newcommand{\kms}{km~s$^{-1}$}

\newcommand {\apgt} {\ {\raise-.5ex\hbox{$\buildrel>\over\sim$}}\ }
\newcommand {\aplt} {\ {\raise-.5ex\hbox{$\buildrel<\over\sim$}}\ }

\title[Sr]
{Enrichment of the Galactic disc with neutron capture elements: Sr 
\thanks{Based on observations collected at OHP observatory, France}
\thanks{Table A1 is only available in electronic form}
}
\author[T.~Mishenina  et al.]
{T.~Mishenina$^{1}$,
 M.~Pignatari$^{2,3,4}$,
T.~Gorbaneva$^{1}$,
S.~Bisterzo$^{5,6,4}$,
C.~Travaglio$^{5,6,4}$,\newauthor
F.-K.~Thielemann$^{7}$,
 C.~Soubiran$^{8}$
 \\
$^{1}$Astronomical Observatory, Odessa National University,         and \\
    Isaac Newton Institute of Chile, Odessa branch,
       Shevchenko Park, 65014, Odessa, Ukraine\\
$^{2}$ E.A. Milne Centre for Astrophysics, Dept of Physics \& Mathematics, University of Hull, HU6 7RX, United Kingdom\\
$^{3}$    Konkoly Thege Miklos ut 15-17, H-1121 Budapest, Hungary\\
$^{4}$ The NuGrid collaboration, http://www.nugridstars.org\\ 
$^{5}$ INFN, Istituto Nazionale Fisica Nucleare, Via Pietro Giuria, 1, 10125, Turin, Italy \\
$^{6}$ B2FH Association, Turin, Italy \\
$^{7}$ Department of Physics, University of Basel, Klingelbergstrabe 82,
        4056 Basel, Switzerland\\
$^{8}$  Laboratoire d'Astrophysique de Bordeaux, 
        Univ. Bordeaux  - CNRS, B18N,  all\'ee Geoffroy Saint-Hilaire, 33615 Pessac, France\\}

\begin{document}

\date{Accepted 2015 xxx. Received 2015 xxx; in original form 2015 xxx}
\pagerange{\pageref{firstpage}--\pageref{lastpage}}
\pubyear{2015}

\maketitle

\label{firstpage}

\begin{abstract}
The enrichment history of heavy neutron-capture elements in the Milky Way disc provides fundamental information about the chemical evolution of our Galaxy and about the stellar sources that made those elements. In this work we give new observational data for Sr, the element at the first neutron-shell closure beyond iron, N=50,
based on the analysis of the high resolution spectra of 276 Galactic disc stars.  
The Sr abundance was derived by comparing the observed and synthetic spectra in the region of the Sr I 4607 \AA~ line, making use of the LTE approximation. 
NLTE corrections lead to an increase of the abundance estimates obtained under LTE, but for these lines they are minor near solar metallicity. The average correction that we find is 0.151 dex. The star that is mostly affected is HD 6582, with a 0.244 dex correction.   
The behavior of the Sr abundance as a function of metallicity is discussed within a stellar nucleosynthesis context, in comparison with the abundance of the heavy neutron-capture elements Ba (Z=56) and Eu (Z=63). The comparison of the observational data with the current GCE models confirm that the $s$-process contributions from Asymptotic Giant Branch stars and from massive stars are the main sources of Sr in the Galactic disc and in the Sun, while different nucleosynthesis sources can explain the high [Sr/Ba] and [Sr/Eu] ratios observed in the early Galaxy.
\end{abstract}

\begin{keywords}
stars: abundances -- stars: late-type -- Galaxy: disc -- Galaxy: evolution
\end{keywords}

\section{Introduction}

The study of the chemical enrichment history of stars in our Galaxy allows to benchmark our understanding in its formation and evolution, and in stellar evolution and nucleosynthesis. Across the evolution of the Galaxy elements have been made by different generations of stars, building up the abundance pattern observed today also in the Sun \citep[e.g.,][]{matteucci:86, timmes:95, goswami:00, gibson:03, kobayashi:11}. Despite their low abundance relative to other metals lighter than iron, heavy elements provide powerful constraints for chemical evolution and other nuclear astrophysics disciplines. According to the established scenario of nucleosynthesis of heavy elements in stars, about half of the abundances beyond iron are due to the slow neutron capture process or $s$-process \citep[e.g.,][and references therein]{kaeppeler:11}, and half to the rapid neutron capture process, or $r$-process \citep[e.g.,][and references therein]{thielemann:17, cowan:19}. 
However, in the last twenty years a growing amount of theoretical and observational works provide the evidence of the existence of other nucleosynthesis processes feeding the production of heavy elements, at least up to the first neutron-magic peak beyond Fe, where elements Sr, Y and Zr are located. Different types of neutrino-driven wind components from forming proto-neutron stars in Core-Collapse Supernovae (CCSNe) have been shown to potentially contribute to the production of these elements, at least in the early Galaxy \citep[e.g.,][]{froehlich:06,farouqi:09,arcones:11,roberts:10,wanajo:11, martinez:14, curtis:19}. \cite{wanajo:11a} discussed as well the nucleosynthesis production of these elements in electron-capture supernovae.

 A number of nucleosynthesis processes needed at low metallicity have been also discussed by \cite{hansen:14}, 
utilizing the approach by \cite{qian:01}. These authors considered neutrino-driven winds in CCSNe as a source of Sr. More recent analyses of production of elements at the Sr peak in metal-poor stars are provided by \cite{hansen:18} and \cite{spite:18}.

In this context, a clear understanding of the production of elements in the Sr-Y-Zr region becomes more complicated compared to the established two-components scenario, where only the $s$-process and the $r$-process are relevant. Many processes need to be taken into account to explain the observed abundances, and their relative relevance may change across the history of the Galaxy.
In the Galactic halo the role of $s$-process production from Asymptotic Giant Branch (AGB) stars in Galactic Chemical Evolution (GCE) is minor, even for elements that are typically classified as $s$-process elements by looking at the abundance distribution in the solar system \citep[e.g.,][]{travaglio:04}. Recent GCE simulations by \cite{bisterzo:14} assign to Sr an $s$-process contribution from AGB stars of 68.9$\pm$5.9 \%, but that contribution is not significant for Sr observed in metal-poor stars. On the other hand, $s$-process in fast rotating metal poor stars could provide a significant contribution to Sr production observed in Galactic halo stars \citep[][]{pignatari:08,frischknecht:16}, and could be marginal for the Sr abundance in the Galactic disc. A study of the Sr/Ba ratio in four halo stars \citep[][]{spite:14} has shown that the abundance pattern of the $s$-process elements is strikingly similar to the theoretical estimates of the $s$-process. The contribution to the $s$-process by rapidly rotating stars \citep{meynet:17,choplin:17,nishimura:17} as the missing component responsible for the relative distribution of the light (Sr) and heavy (Ba) neutron-capture elements has been studied by adopting a stochastic chemical evolution model \citep[][]{cescutti:15b}.

\cite{travaglio:04} found that in the solar abundances there is a component missing between Sr and Xe, not explained by the traditional $s$-process and $r$-process scenario. They called that component Lighter Element Primary Process or LEPP, and associated to the Sr-rich signature observed in a large fraction of metal-poor stars. This result is still controversial \citep[see e.g.,][]{honda:04,honda:07,montes:07,trippella:16,cristallo:15}. Travaglio et al. results were not taking into account all the zoo of processes possibly feeding at least the Sr-Y-Zr peak, and it is plausible that some of them will be relevant for GCE.
Sr is made from the weak $s$-process in massive stars \citep[e.g.,][]{raiteri:91a,raiteri:91b,pignatari:10,the:07,pignatari:16a} and in massive AGB stars ($>$ 4 M$_\odot$) \citep[e.g.,][]{karakas:14,cristallo:15,pignatari:16b}. \cite{travaglio:04} estimated the contribution by the weak $s$-process to be 9\% of the solar Sr. The contribution from massive AGB stars changes between 9\% \citep[][]{travaglio:04} to 1.35\% by \citep[][]{bisterzo:14}. There is not a clear estimate of the errors associated to these contributions, where both nuclear and stellar model uncertainties are consistently taken into account \citep[e.g.,][]{pignatari:16b}. 
The $r$-process, together with all of these nucleosynthesis processes, made the remaining part of Sr that was not created by the $s$- process.
However, also the origin of the $r$-process elements with A $>$ 56 remains controversial. At least four sources have been proposed, namely: 1) the neutrino-induced winds from supernovae \citep{woosley:94,takahashi:94}; 2) the neutron-rich matter ejected from coalescencing neutron stars \citep{freiburghaus:99,thielemann:17} (see further references in the latter review);  3) the winds from the black hole-neutron stars mergers \citep{surman:08}; and 4) polar jet ejecta from magneto-rotational supernove \citep{winteler:12,nishimura:15,nishimura:17}.  

In the last years the intermediate-neutron capture process or $i$-process \citep[][]{cowan:77} has been shown to be active since the first stages of the evolution of the Galaxy, possibly explaining anomalous abundance patterns observed in old metal-poor stars \citep[e.g.,][]{dardelet:14, hampel:16, roederer:16, clarkson:18}, in younger objects in the Galactic disc and in open clusters \citep[][]{herwig:11, mishenina:15, dorazi:17} and in presolar grains \citep[][]{fujiya:13,liu:14}.

Also alternative sources have been introduced in several papers \citep{travaglio:04,qian:08}. For example, the role of neutron star mergers in the chemical evolution of the Galactic halo and the $r$-process production of Sr, Zr, and Ba - complemented by an $s$-process production from spinstars was presented in \cite{cescutti:15a}.  Both neutron star mergers and supernova scenarios might have contribute in producing Eu, and observations at low metallicity allow to identify two components of $r$-process nucleosynthesis \citep[e.g.,][]{wehmeyer:15}.
Indeed, theoretical $r$-process estimates can be tested directly with Galactic Archaeology, by looking at the composition of stars formed 
with insufficiently mixed matter, and enriched with heavy elements resulting from one or few early $r$-process events \cite[e.g.][]{aoki:07,sneden:08,roederer:10}.

The Sr abundance was studied in 156 stars of the Galactic disc in a recent paper by \cite{battistini:16}. The authors concluded that the $s$-process is responsible for the main contribution in the enrichment of Sr,  with an additional contribution from a non-classical $r$-process at low metallicities. In the thin disc the trends of [El/Fe] vs. [Fe/H] are flatter, which is due to the fact that the main production from the $s$-process is balanced by Fe production from type Ia supernovae.
With metallicities in the range from -1 $<$ [Fe/H] $<$ 0.3 dex the contributions to neutron capture elements by all mentioned processes are different, and they change in the course of the Galaxy evolution.
In previous studies we have determined the abundances of a number of neutron-capture elements for more than 250 stars \citep[][]{mishenina:13}. 
Here we extend our study with information on Sr, and we provide a comparative analysis of the abundances of elements that in the Galactic disc are
mostly made by the $s$-process (Sr and Y at the neutron shell closure N=50, and Ba and La at N=82) $s$-process elements in relation to europium (Eu), produced by $r$-process. 

The paper is organized as follow. 
The observations and selection of stars plus the definition of the main stellar 
parameters are described in \S \ref{sec: stellar param}. 
 The abundance determinations  and the error analysis are presented in 
\S \ref{sec: abundance determination}. 
The results and comparison with other data 
as well as the application of the results to the theory of nucleosynthesis and the chemical evolution of the Galaxy are reported in \S \ref{sec: result, gce}. 
Conclusions are drawn in \S \ref{sec: conclusions}.

\section{Observations and atmospheric parameters}
\label{sec: stellar param}

Most observations used here were previously analysed in our paper on n-capture elements \citep{mishenina:13}.  The spectra were obtained using the 1.93 m telescope at Observatoire de Haute-Provence (OHP, France) equipped with the echelle type spectrographs ELODIE (R = 42000 ) for the wavelengths range 4400 -- 6800 \AA~  and signal to noise S/N more than 100. Our starting sample includes 276 stars like in \cite{mishenina:13}. Also for those stars we have searched additional spectra in the OHP spectroscopic archive \citep{moultaka:04}, from the SOPHIE spectrograph which cover a similar wavelength range at a resolution of R= 75000. The primary processing of spectra was carried out immediately during observations \citep{katz:98}. Further spectra processing such as the continuum placement, line depth and equivalent width (EW) measurements, etc., was conducted using the 
DECH20 software package by \cite{galazut:92}.

This paper belongs to a set of studies of abundances in stars in the galactic disc \citep{mishenina:04, mishenina:08, mishenina:13}. 
We use the same stellar parameters derived for stars in our sample. To estimate the effective temperatures \Teff\ , we used one and the same approach for 267 dwarfs in our sample; in so doing, for better control we have applied the far-wing fitting of the H$_\alpha$ line profiles for nine stars with metallicities below --0.6 dex, and that turned out to be more suitable. Since the  far-wings of H$_\alpha$ are independent from gravity, metallicity and convection of the atmosphere model \citep{gratton:96}, and also  to avoid uncertainties in the calibrations which  were constructed in the range of --0.5 $<$ [Fe/H] $<$ +0.5 and used by us for a large part of dwarfs. Effective temperatures \Teff\ were determined by the calibration of line-depth ratios for spectral line pairs with significantly different low-level excitation potential applying the technique introduced and developed by \cite{kovtyukh:03}. The mean random error of each single calibration was 60–-70 K (it ranged from 40-–45 K to 90-–95 K for the most and least accurate calibrations, respectively). The usage of about 70 -– 100 calibrations enabled us to reduce the uncertainty down to 5-–7 K (for the spectra with S/N ratio of 100-−150). It has been shown that 105 calibrations are essentially independent of micro-turbulence, departures from LTE (Local Thermodynamic Equilibrium), elemental abundances, rotational parameters or any other individual stellar properties. The estimated accuracy of the method varied within the range from 5 to 45 K for the dwarfs with [Fe/H] $\ge$ -0.5. For most of metal-poor stars of the sample, \Teff was estimated by the far-wing fitting of the H$_\alpha$ line profiles \citep{mishenina:01}. We have proved in \cite{mishenina:04} that the temperature scales adopted in \citep{mishenina:01} and \citep{kovtyukh:03} are well consistent.

Surface gravities \logg\ were computed by the ionization balance, implying that the iron abundances obtained from the neutral iron Fe~{\sc i} and ionized iron Fe~{\sc ii} lines were similar. The two most--commonly used techniques for the surface gravity determination are the ionization balance of neutral and ionized species and the fundamental relation expressing the gravity as a function of the mass, temperature and bolometric absolute magnitude deduced from the parallax. A detailed study of surface gravities derived by different methods was performed by \cite{allende:99}, who reported that astrometric and spectroscopic (iron ionization balance) gravities were in good agreement within the metallicity range of -−1.0 $<$ [Fe/H] $<$ +0.3. In our earlier paper \citep{mishenina:04}, we compared the adapted surface gravities to those determined astrometrically by \cite{allende:99}; the resulting mean difference and standard deviation were -−0.01 and 0.15, respectively, for 39 common stars. This is consistent with an accuracy of 0.1 dex of our spectroscopic gravity determinations. Moreover, in each of our studies, we have been analysing the correlation between our estimates of chemical abundances and stellar parameters to justify the correctness of the latter. 

The adopted value of the metallicity [Fe/H] was calculated using the iron abundance obtained from the Fe~{\sc I} lines. As is known \citep[e.g.][]{thevenin:99,shchukina:01,mashonkina:11,bergemann:12}, the lines of neutral iron are influenced by the deviations from the LTE in solar and stellar spectra, and hence, these deviations also affect the iron abundances determined from those lines. However, within the temperature and metallicity ranges of our target stars, the NLTE corrections do not exceed 0.1 dex
 \citep[see, e.g.][]{mashonkina:11}.

The microturbulent velocity \Vt\ was derived considering that the iron abundance obtained from a given Fe~{\sc i} line is not correlated with the equivalent width EW of that line.

The obtained parameter values and their comparison with the results of other authors are reported in  \cite{mishenina:04, mishenina:08, mishenina:13}. The accuracy of our parameter determination is estimated to be: 
$\Delta$\Teff = $\pm100~K$, $\Delta$\logg = $\pm0.2$dex,  $\Delta$\Vt = $\pm0.2$\kms, $\Delta$[Fe/H] = $\pm0.1$dex. 
In this study, we have compared the adopted parameters with those obtained recently by \cite{battistini:16} and \cite{delgado:17} who reported the Sr abundances estimated in the LTE approximation using the same Sr I line as in this study. In particular, our goal was to assess the \Teff\ scale in our study, which is essential for the Sr abundance determinations. The results of the comparison for individual stars are given in Table \ref{comp_batt} while Table \ref{ncap} presents the mean differences and errors (standard deviations) in the parameter values for the common set of target stars in various papers. In these Tables, we have also provided the results of the comparison of our Sr data with those obtained earlier \citep{reddy:03,mashonkina:01,brewer:06}; note, the Sr II line was used in the last two studies. 

\begin{table*}
\begin{center}
\caption{Parameters of our target stars and comparison with \citet{battistini:16,delgado:17,mashonkina:01, reddy:03,brewer:06} for common stars.}
\label{comp_batt}
\begin{tabular}{lcccccccccc}
\hline
 HD & \Teff, K & \logg & [Fe/H] &  HD & \Teff, K  &  \logg  & [Fe/H]& $\Delta$ \Teff, K& $\Delta$ \logg 
	   & $\Delta$ [Fe/H]   \\
\hline
      & our    &    &     &  &\citet{battistini:16}        &     &     &     &      &      \\ 
\hline 
 8648	&5841	&4.3	&0.22	&8648	&5790	&4.2	&0.12	& 51	& 0.1	& 0.1	     \\
22879	&5972	&4.5	&-0.77	&22879	&5825	&4.42	&-0.91	& 145	& 0.08	& 0.1	      \\
30495	&5790	&4.5	&0.02	&30495	&5820	&4.4	&-0.05	&-30	&0.1	&0.07	      \\
64606	&5188	&4.4	&-0.91	&64606	&5250	&4.2	&-0.91	&-62	&0.2	&0	     \\
64815	&5763	&3.9	&-0.35	&64815	&5864	&4	&-0.33	&-101	&-0.1	&-0.02	     \\
135204	&5200	&4.4	&-0.19	&135204	&5413	&4	&-0.16	&-213	&0.4	&-0.03	     \\
152391	&5322	&4.5	&-0.08	&152391	&5495	&4.3	&-0.08	&-173	&0.2	&0	     \\
157089	&5915	&4.3	&-0.5	&157089	&5785	&4	&-0.56	&130	&0.3	&-0.06	     \\
159482	&5760	&4.3	&-0.81	&159482	&5620	&4.1	&-0.89	&140	&0.2	&0.08	      \\
159909		&5671	&4.3	&0.03	&159909	&5749	&4.1	&0.06	&-78	&0.2	&-0.03	     \\
165401	&5794	&4.5	&-0.4	&165401	&5877	&4.3	&-0.36	&-83	&0.2	&-0.04	     \\    
178428	&5656	&4.2	&0.15	&178428	&5695	&4.4	&0.14	&-39	&-0.2	&0.01	      \\   
187897	&5944	&4.5	&0.12	&187897	&5887	&4.3	&0.08	&57	&0.2	&0.04	     \\    
190360	&5572	&4.5	&0.26	&190360	&5606	&4.4	&0.12	&-34	&0.1	&0.14	      \\   
199960	&6023	&4.4	&0.33	&199960	&5878	&4.2	&0.23	&145	&0.2	&0.1	      \\   
201891	&5973	&4.3	&-1.08	&201891	&5850	&4.4	&-0.96	&123	&-0.1	&-0.12	      \\ 
217014	&5858	&4.4	&0.24	&217014	&5763	&4.3	&0.17	&95	&0.1	&0.07	     \\     
\hline
      &  our    &    &     &  &  \citet{delgado:17}      &     &     &     &      &      \\ 
\hline
4307  	&5889	&4.0	&-0.18	&4307	&5840	&4.13	&-0.21	&49	&-0.13	&0.03       \\    
14374	&5449	&4.3	&-0.09	&14374	&5375	&4.42	&-0.07	&74	&-0.12	&-0.03        \\   
22049	&5084	&4.4	&-0.15	&22049	&5049	&4.45	&-0.15	&35	&-0.05	&0.0        \\    
22879	&5972	&4.5	&-0.77	&22879	&5949	&4.68	&-0.79	&23	&-0.18       &0.02       \\   
38858	&5776	&4.3	&-0.23	&38858	&5719	&4.49	&-0.23	&57	&-0.19	&0.00       \\   
76151	&5776	&4.4	&0.05	        &76151	&5781	&4.44&0.12	        &-5	&-0.04	&-0.07        \\ 
125184	&5695	&4.3	& 0.31	&125184	&5660	&4.11	&0.27	        &35	&0.19 	&0.04        \\   
146233	&5799	&4.4	& 0.01	&146233	&5810	&4.46	&0.05  	&-11	&-0.06	&-0.04        \\   
161098	&5617	&4.3	&-0.27	&161098	&5574	&4.49	&-0.26	&43	&-0.19	&-0.01        \\   
199960	&5878	&4.2	&0.23	        &199960	&5928	&4.42	&0.27 	&-50	&-0.22	&-0.04        \\ 
210752	&6014	&4.6	&-0.53	&210752	&5970	&4.52	&-0.55	&44	&0.08 	&0.02       \\    
\hline 
 &  our    &    &     &  & \citet{mashonkina:01}       &     &     &     &      &      \\
\hline
4614	&  5965	&4.4	&-0.24	&4614 	&  5940	&4.33	&-0.3	 &25	 & 0.07	& 0.06   \\
22879 &	5972	&4.5	&-0.77	&22879	&  5870	&4.27	&-0.86	&102	& 0.23&	 0.09  \\
55575	&  5949	&4.3	&-0.31	&55575	&  5890	&4.25	&-0.36	&59	  &0.05	& 0.05   \\
64606	&  5250	&4.2	&-0.91	&64606	&  5320	&4.54	&-0.89	&-70	 &-0.34&	-0.02  \\
65583	&  5373	&4.6	&-0.67	&65583	&  5320	&4.55	&-0.73	&53	  &0.05 &	0.06  \\
68017	&  5651	&4.2	&-0.42	&68017	&  5630	&4.45	&-0.40	&21	 &-0.25	&-0.02  \\
109358&	5897	&4.2	&-0.18	&109358	&5860	&4.36	&-0.21	 &37	 &-0.16	&0.03  \\
112758&	5203	&4.2	&-0.56	&112758	&5240	&4.62	&-0.43	&-37	 &-0.42	&-0.13  \\
114710&	5954	&4.3	& 0.07	&114710	&6000	&4.30	&-0.03	&-46	 & 0.0	 &  0.1   \\
117176&	5611	&4.0  &-0.03	&117176	&5480	&3.83	&-0.11	&131	  &0.17	&0.08  \\
126053&	5728	&4.2	&-0.32	&126053	&5690	&4.45	&-0.35	&38	   &-0.25	&0.03 \\
144579&	5294	&4.1	&-0.70	&144579	&5330	&4.59	&-0.69	&-36	 &-0.49	&-0.01 \\
168009&	5826	&4.1	&-0.01	&168009	&5785	&4.23	&-0.03	&41	  &-0.13	& 0.02 \\
176377&	5901	&4.4	&-0.17	&176377	&5860	&4.43	&-0.27	&41	  &-0.03	& 0.1  \\
\hline
 &  our    &    &     & &  \citet{reddy:03}       &     &     &     &      &      \\
\hline
11007& 	5980 &	4	& -0.2&  11007&	5850&	4	&-0.31 & 130&	0&	0.11 \\
42618&	5787& 	4.5&	-0.07& 42618&	5653&	4.58&	-0.16& 134&	-0.08&	0.09  \\
45067&	6058&	4&	-0.02& 45067&	5946&	3.99&	-0.12& 112&	0.01&	0.1 \\
71148&	5850& 	4.2&	0& 71148&	5703&	4.46&	-0.08 &147&	-0.26&	0.08   \\
126053&	5728&	4.2&	-0.32& 126053&	5597&	4.44&	-0.41 &131&	-0.24&	0.09\\
186408&	5803&	4.2&	0.09&  186408&	5670&	4.32&	0& 133	&-0.12&	0.09 \\
206860&	5927&	4.6&	-0.07&206860&	5820&	4.48&	-0.12& 107&	0.12&	0.05\\
\hline 
  &  our    &    &     &  & \citet{brewer:06}       &     &     &     &      &      \\
\hline
25665& 	4967&	4.7&	0.01&	25665&	4870&	4.4	&-0.012&  97	&0.3&	0.022 \\
53927&	4860&	4.64&	-0.22&	53927&	4960&	4.6&	-0.385&  -100&	0.04&	0.165 \\
159062&	5414&	4.3&	-0.4&	159062&	5260&	4.45&	-0.507 &154	&-0.15&	0.107  \\
168009&	5826&	4.1&	-0.01&	168009&	5720&	4.2&	-0.07&  106&	-0.1&	0.06\\
\hline                                                                                         
\end{tabular} 
\end{center}  
\end{table*}

\begin{table*}
\begin{center}
\caption[]{Comparison of our parameters and Sr abundance determinations with 
the results of other authors for the $n$ stars shared with our stellar sample. 
}
\label{ncap}
\begin{tabular}{cccccc}
\hline
 Reference & $\Delta$(\Teff) & $\Delta$(\logg) & $\Delta$([Fe/H]) & $\Delta$([Sr/Fe]) & n \\
\hline
Battistini \& Bensby& 4 & 0.13 & 0.03 & --0.01 & 17 (1)\\
 2016 &$\pm$116 & $\pm$0.15 & $\pm$0.07 & -- &  \\
Delgado Mena et al.& 27 & -0.08 & -0.01 & --0.05 & 12 \\
 2017&$\pm$36 & $\pm$0.13 & $\pm$0.03 & $\pm$0.09 &  \\
Mashonkina \& Gehren& 26 & -0.10 & 0.03 & 0.02 & 14  \\
 2001  &$\pm$56& $\pm$0.21 & $\pm$0.06 & $\pm$0.10 &  \\
Reddy et al. & 127 & -0.08 & 0.09 & --0.03 & 7  \\
 2003 &$\pm$13 & $\pm$0.14 & $\pm$0.02 & $\pm$0.08 &  \\
Brewer \& Carney  & 64 & 0.02 & 0.09 & -0.21 & 4 \\
 2006  &$\pm$112 & $\pm$0.20 & $\pm$0.06 &  $\pm$0.22 &   \\
\hline
\end{tabular}
\end{center}
\end{table*}

We find the concordance between our data and those by \cite{battistini:16} within the stated error definitions, except for \Teff\ for the stars HD 135204, 152391, 157089, 159482, 199960, 201891 and 
for \logg~ for the star HD 135204. At that the average difference values of $<$ $\Delta$ \Teff\ $>$,  $<$ $\Delta$ \logg $>$, $<$ $\Delta$ [Fe/H] $>$ are equal to --4  $\pm 116 $,  --0.13 $\pm0.15$, --0.03  $\pm0.07$, respectively.
Matching our results with those of  \cite{delgado:17} we obtained the average values $<$ $\Delta$ \Teff\ $>$ = 
27 $\pm 36 $,  $<$ $\Delta$ \logg $>$ = --0.08 $\pm0.13$, $<$ $\Delta$ [Fe/H] $>$ = --0.01 $\pm0.03$,
which  show a good agreement between themselves.

Earlier, we carried out the kinematic classification of the thin and thick disc stars, as well as of the Hercules stream stars \citep{mishenina:04}, based on the Hipparcos (ESA 1997) parallaxes and proper motions combined with radial velocities measured by the cross-correlation of the ELODIE spectra (with an accuracy better than 100 m s$^{-1}$). We have not updated our classification with respect to the latest astrometric data from the {\it Gaia} Data Release 2 \citep{GDR2:18} either due to the fact that many stars of our sample are too bright to be measured by {\it Gaia} or that the relevant astrometric errors are equivalent to those of the Hipparcos observations. The classification is based on the (U, V, W) velocities with respect to the Sun with typical errors of 1 km s$^{-1}$. Having assumed that our sample represents three populations of stars in the solar vicinity, such as those of the thin and thick disc, as well as the Hercules stream group, we have computed the probability of each star's membership in either of these populations. In these computations, we have adapted the velocity ellipsoids determined by \cite{soubiran:03}. A star is considered to belong to a certain population if the probability was higher than 70\%. Application of this criterion implies that there are a number of stars with intermediate kinematics which cannot be classified.

\section{Determination of Sr abundances}
\label{sec: abundance determination}

\begin{figure}
\begin{tabular}{c}
\includegraphics[width=8cm]{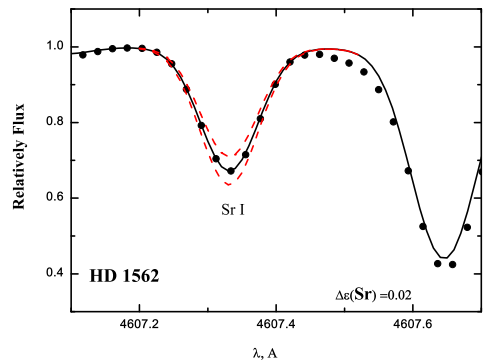}\\
\end{tabular}
\caption{Observed (dots) and calculated (solid and dashed lines) spectra in the region 
of Sr I line  for HD 1562, the change in the Sr abundance is 0.02 dex.}
\label{sr_prof}
\end{figure}

The determination of the Sr abundance was obtained with the new version of the STARSP LTE spectral synthesis code \citep{tsymbal:96} from the Sr I line 4607 \AA~ using the stellar models \citep{castelli:04}. A comparison of synthetic and observed spectra for the Sr line is shown in Fig. \ref{sr_prof}. 

The Sr abundance was determined by differential analysis relative to solar one. Solar abundances were calculated using the solar profiles, measured in the spectra of the Moon and asteroids; they were also estimated using the SOPHIE spectrograph and the oscillator strengths log\,gf adopted from the VALD database \citep{kupka:99}. 
Our approved LTE solar Sr abundance is log A(Sr)$_\odot$ = 2.74$\pm$0.03 {in} comparison to 
  2.87 $\pm$0.07   \citep{asplund:09}, 2.83 $\pm$0.06 \citep{grevesse:15}, and 2.78 \citep{delgado:17}.
It should be emphasized that in \cite{battistini:16} the values of the solar Sr abundance determined from the line 4607 \AA~ in the spectra of reflected sunlight obtained from different spectrographs with various resolutions are given. These values noticeably different (about 0.2 dex), ranging from log A(Sr)$_\odot$(MIKE) = 2.69 to log A(Sr)$_\odot$(FEROS)= 2.92,  where log A(H) = 12.0. This is important to keep in mind and take into account when determining the content of elements relative to solar one, since it may be generate a systematic shift of observational data. 
The departures from LTE and their effect on the determination of the Sr abundances for stars with different metallicities have been investigated in a number of papers \cite[e.g.][]{belyakova:97, mashonkina:01, andrievsky:11}, wherein the Sr II lines were analysed. An NLTE analysis of the Sr I and Sr II lines in the spectra of late-type stars was performed by \cite{bergemann:12}. The model of the Sr atom was constructed using the atomic data available in the Hannover and NIST databases. The neutral atom was represented by 141 levels; the singly-ionized atom included 49 levels. The described model of Sr was similar to that one created by \cite{andrievsky:11} with regard to the term structure and the number of dipole--permitted transitions of Sr II, but unlike the latter it factored in the effect of deviations from LTE on the neutral Sr line (for more details see \cite{bergemann:12}). A grid of the NLTE abundance corrections for Sr I and Sr II lines was presented in \citep{bergemann:12}. The NLTE corrections for the Sr I line 4607 \AA~ reported in \citep{bergemann:12} for dwarfs varied within the range of 0.10 -- 0.23 dex at [Fe/H] $>$ --0.8 dex, depending on the temperature and metallicity of star. Using the data of \cite{bergemann:12}, we have interpolated the values of the NLTE corrections for the Sr I line 4607 \AA~ for our target stars. The NLTE Sr correction for the Sun is 0.10 dex. For metal poor stars it is more suitable to use the Sr II lines which have smaller NLTE corrections, not exceeding 0.2 dex \citep{andrievsky:11} or 
close to 0.05 dex \citep{hansen:13}.

\par
The obtained  LTE Sr abundances,  the NLTE corrections from \cite{bergemann:12}, the NLTE Ba and LTE Eu abundance,  and stellar parameters \citep{mishenina:13} are given in Table A1
which is available on-line.

Fig. \ref{sr_LTE_NLTE} presents our observations  and a comparison with GCE predictions by \cite{bisterzo:14} and \cite{travaglio:04}, and also the interpolated NLTE corrections from \cite{bergemann:12}. Fig. \ref{comp_obs} shows a comparison between our data and those of \cite{battistini:16} and \cite{delgado:17}, with GCE model by \cite{bisterzo:14}.

\subsection{Errors in abundance determinations}

\begin{figure}
\begin{tabular}{c}
\includegraphics[width=8cm]{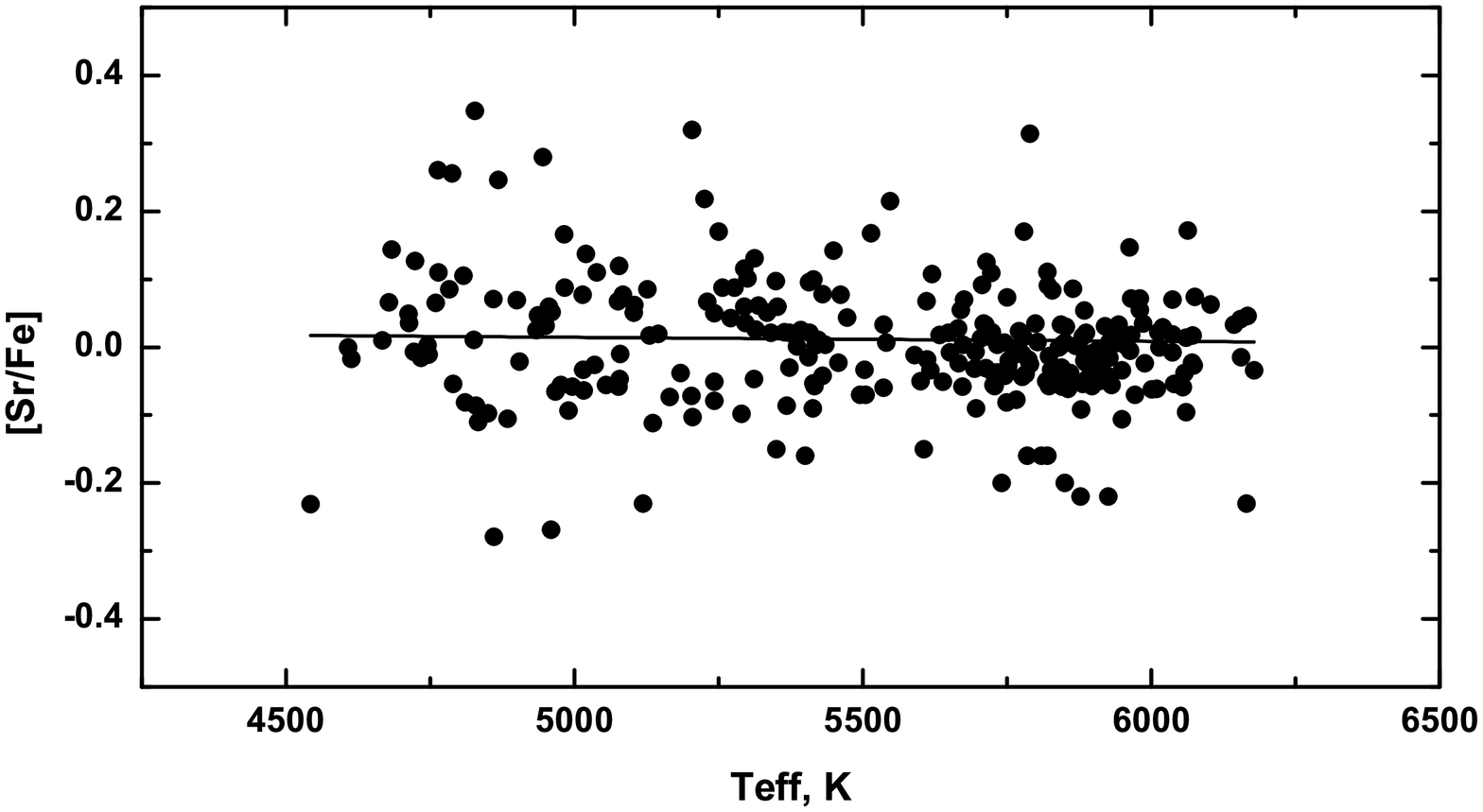}\\
\end{tabular}
\caption{Dependence of  [Sr/Fe] vs. \Teff.}
\label{sr_teff_}
\end{figure}

\begin{figure}
\begin{tabular}{c}
\includegraphics[width=8cm]{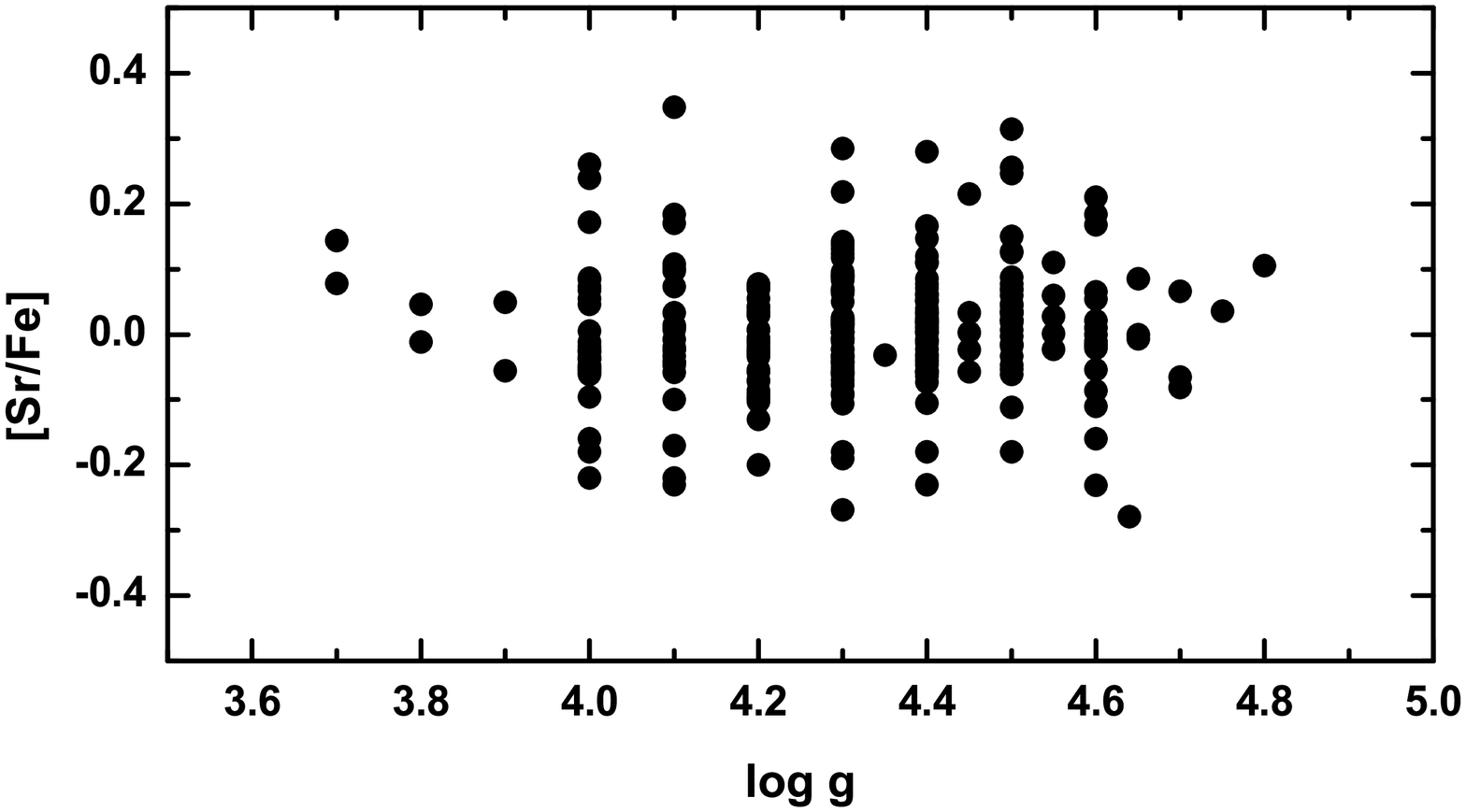}\\
\end{tabular}
\caption{Dependence of  [Sr/Fe] vs. \logg.}
\label{sr_logg_}
\end{figure}

To determine the systematic errors in the elemental abundances, resulting from 
uncertainties in the atmospheric parameters, we derived the elemental
abundance of two stars HD216259 (\Teff\ = 4833 K, \logg\ = 4.60, \Vt = 0.5 km/s, [Fe/H] = -0.55) and HD9826 (\Teff\ = 6074 K,  \logg\ = 4.00, \Vt = 1.3 km/s, [Fe/H] = 0.10) for several models with modified parameters 
($\Delta$\Teff $= \pm100~K$, $\Delta$\logg $= \pm0.2$, $\Delta$\Vt $= \pm0.1$). 
The abundance variations with the modified parameters and the fitting errors
for the computed and observed spectral line profiles (0.02 dex), are given in 
Table \ref{errors}. The maximum contribution to the error is introduced by \Teff.
Total errors due to parameter uncertainties and the measured spectra varies from 0.12 dex for the hot and to 0.06 -- 0.17 dex for the cool  stars.
The dependence of the Sr abundance on stellar parameters (\Teff\ and \logg) is presented in Figs. 
\ref{sr_teff_} and \ref{sr_logg_}. No trend of [Sr/Fe] vs. \Teff\ and \logg\ is observed.

\begin{table*}
\caption{Abundance errors due to atmospheric parameter uncertainties as examples of stars 
with different values of stellar parameters: HD216259 (4833,4.60,0.5,-0.55) and HD9826 (6074,4.00,1.3,0.10).}
\label{errors}
\begin{tabular}{llcccccccc}
\hline
& & HD216259  && & &HD9826 && & \\
 AN & El  & $\Delta$ \Teff+  & $\Delta$ \logg+ & $\Delta$ \Vt+ & tot+ &$\Delta$ \Teff+  & $\Delta$ \logg+ & $\Delta$ \Vt+ & tot+\\
\hline
38	&SrI	&0.15 &--0.07  &--0.04	&0.17	&0.12	&0.00  &--0.02	&0.12	 \\ 
\hline                             
\end{tabular}
\end{table*}

We compare our LTE Sr abundances with the ones obtained by \cite{battistini:16} in LTE assumption who used the same Sr I line 4607 \AA~ as in our case. We have only one star in common (HD 64606) with that work, for which Sr abundance is provided. The difference in the Sr abundance of HD 64606  is consistent within 0.01 dex. 
The mean values of the difference between our LTE definitions and those of \citep{delgado:17} is equal to --0.05 $\pm0.09$, confirming the overall agreement with our determinations. For five stars (HD 22049, HD 22879, HD 38858, HD 125184, HD 161098) the individual differences are larger than 0.05 dex, as highlighted in Fig.~\ref{comp_obs}.  
As can be seen from the figures, a significant scatter in [Sr/Fe] ratio is observed.  In our stellar sample, we obtain an observed range -0.28 $\lesssim$ [Sr/Fe] $\lesssim$ 0.34 for thin disc stars, which is higher than what the range measured for more metal-poor thin-disc stars (from -0.03 $\lesssim$ [Sr/Fe] $\lesssim$ 0.26 dex). This is likely due to a smaller sample of metal-poor stars, providing a less meaningful comparison. We have good agreement with results by \cite{delgado:17} for thin-disc stars with solar-like metallicity (-0.19 $\lesssim$ [Sr/Fe] $\lesssim$ 0.29 dex), while they obtain a larger scatter for Sr abundances in metal-poor stars (-0.36 $\lesssim$ [Sr/Fe] $\lesssim$ 0.40 dex). Greater variation of Sr abundances is shown by \cite{battistini:16}  (-0.37 $\lesssim$ [Sr/Fe] $\lesssim$ 0.54 dex). Their results are obtained in the LTE approximation, but the NLTE corrections are positive, meaning that NLTE corrections will not improve  the situation for moderately metal-poor stars. Taking into account observational data for [Sr/Fe] and their uncertainties, the observed dispersion of [Sr/Fe] is larger than the provided errors: $\pm$ 0.15 dex for our determinations, consistently with \cite{battistini:16}, and from 0.01 to 0.46 dex for the stellar data by \cite{delgado:17}. 
Concerning the [Sr/Fe] trend with respect to [Fe/H], based on our data for thin and thick discs we did not find any significant trend (slope -0.00379$\pm$0.02427). \cite{battistini:16} reported instead a mild increasing [Sr/Fe] abundance ratios with decreasing metallicity: [Sr/Fe] $\approx$ -0.2 for solar metallicity stars,  increasing to [Sr/Fe] $\approx$ 0 at [Fe/H] $\approx$ -−1. 
Taking into account the large [Sr/Fe] scatter observed in the metal-poor stars, giants and dwarfs \citep[e.g.][]{burris:00, brewer:06}, over a metallicity range -2.0 $<$ [Fe/H] $<$ 0.2, the trend seems to be on average solar \citep[e.g.,][]{brewer:06}. \cite{ishigaki:13} observed thick disc and halo stars, also founding solar [Sr/Fe] ratios for [Fe/H] $<$ -1. While there may be contradictory results concerning the observational trend of [Sr/Fe], especially for metal-poor stars  in the thick disc and in the halo, a significant real [Sr/Fe] dispersion beyond the observational error is a consistent result obtained from all authors.

As shown in the works \citep[e.g.][]{belyakova:97,mashonkina:01,andrievsky:11,bergemann:12} the NLTE corrections for the lines of neutral and ionized strontium depend on the stellar parameters (\Teff, \logg,  and [Fe/H]). And, specifically, the dependence on metallicity affects the estimates of the strontium abundance in various Galactic substructures, primarily the halo and the disc, which differ in this parameter. For stars of lowest metallicity (halo), the Sr II line usually used to determine the strontium abundance, for which NLTE corrections are small \citep[e.g.][]{andrievsky:11}. The use of the Sr I lines requires the NLTE corrections from 0.05 to 0.5 \citep[e.g.][]{bergemann:12}, depending on the metallicity. As we show in Fig. \ref{sr_LTE_NLTE}, in our considered metallicity range, the average value of NLTE corrections  is 0.151 dex, for stars of the thin disc, and for stars of the thick disc they change from 0.137 to 0.244 dex and there is the dependence on  metallicity. The considering of the Sr abundance behaviour in the thin and thick disc stars has shown that the NLTE deviations change more the trend for the thick disc stars  than for the thin disc stars. However, the scatter of the Sr abundance on all metallicity of the disc (and the Galaxy) does not allow a fine comparison of the Sr abundance in various substructures of the disc with predictions of models of Galactic evolution.

\section{Results and Comparison with GCE Models}
\label{sec: result, gce}

Element abundances measured in stars are an ideal yardstick for nucleosynthesis predictions and their effect on stellar and galactic evolution. The solar system abundance of the element Sr is dominated by the $s$-process contribution to $^{88}$Sr (82.6\% of the solar Sr) by AGB stars.
$^{86}$Sr and $^{87}$Sr (9.9\% and 7.0\% of the solar Sr, respectively) are $s$-only isotopes \citep[][]{kaeppeler:11}. 
Finally, the rarest Sr isotope is $^{84}$Sr (0.56\% of the solar Sr), that is a product of p-process in stars \citep[e.g.,][and references therein]{rauscher:13,pignatari:16b,travaglio:18}.

\begin{figure*}
\begin{tabular}{c}
\includegraphics[width=16cm]{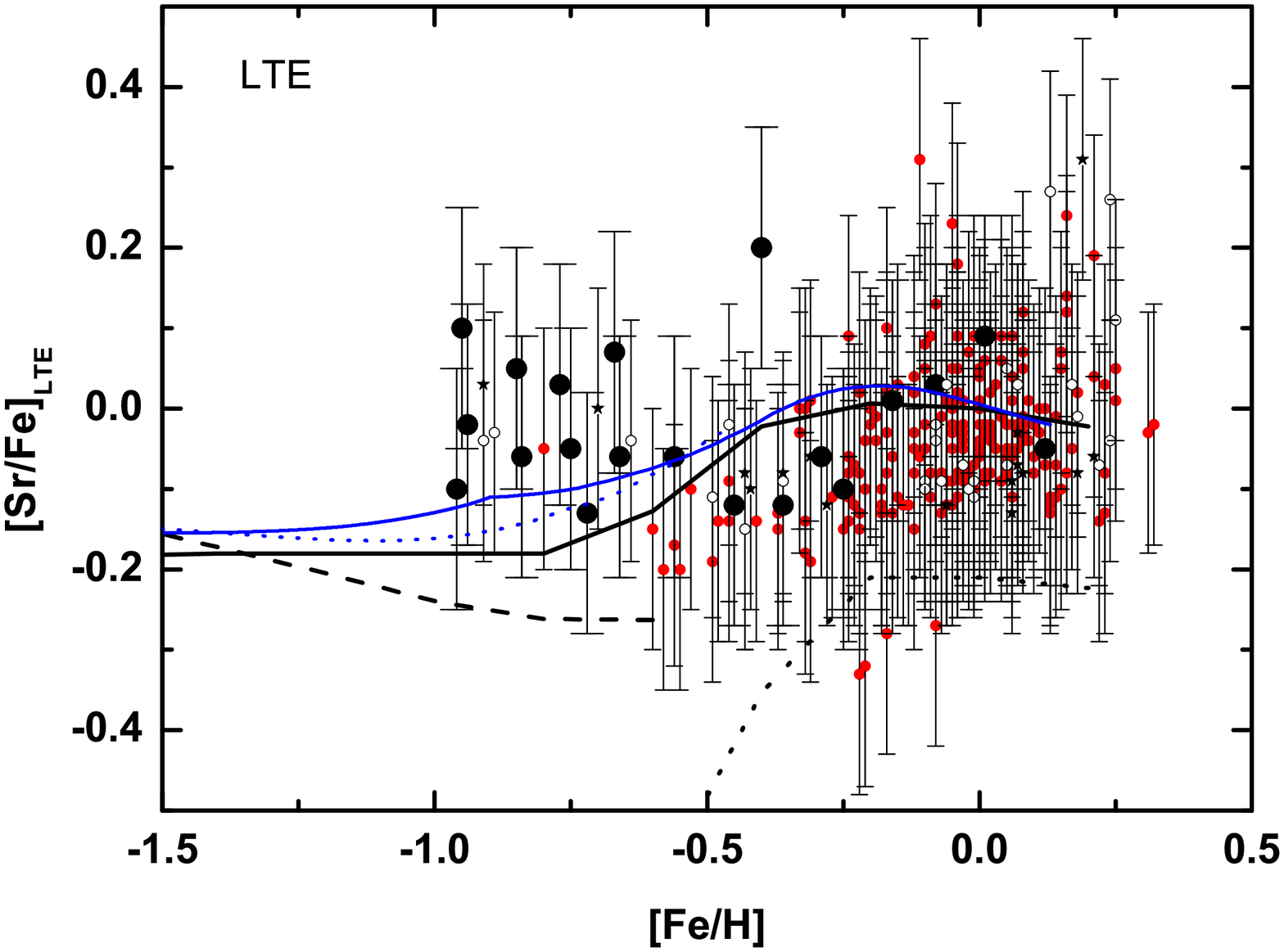}\\ 
\includegraphics[width=16cm]{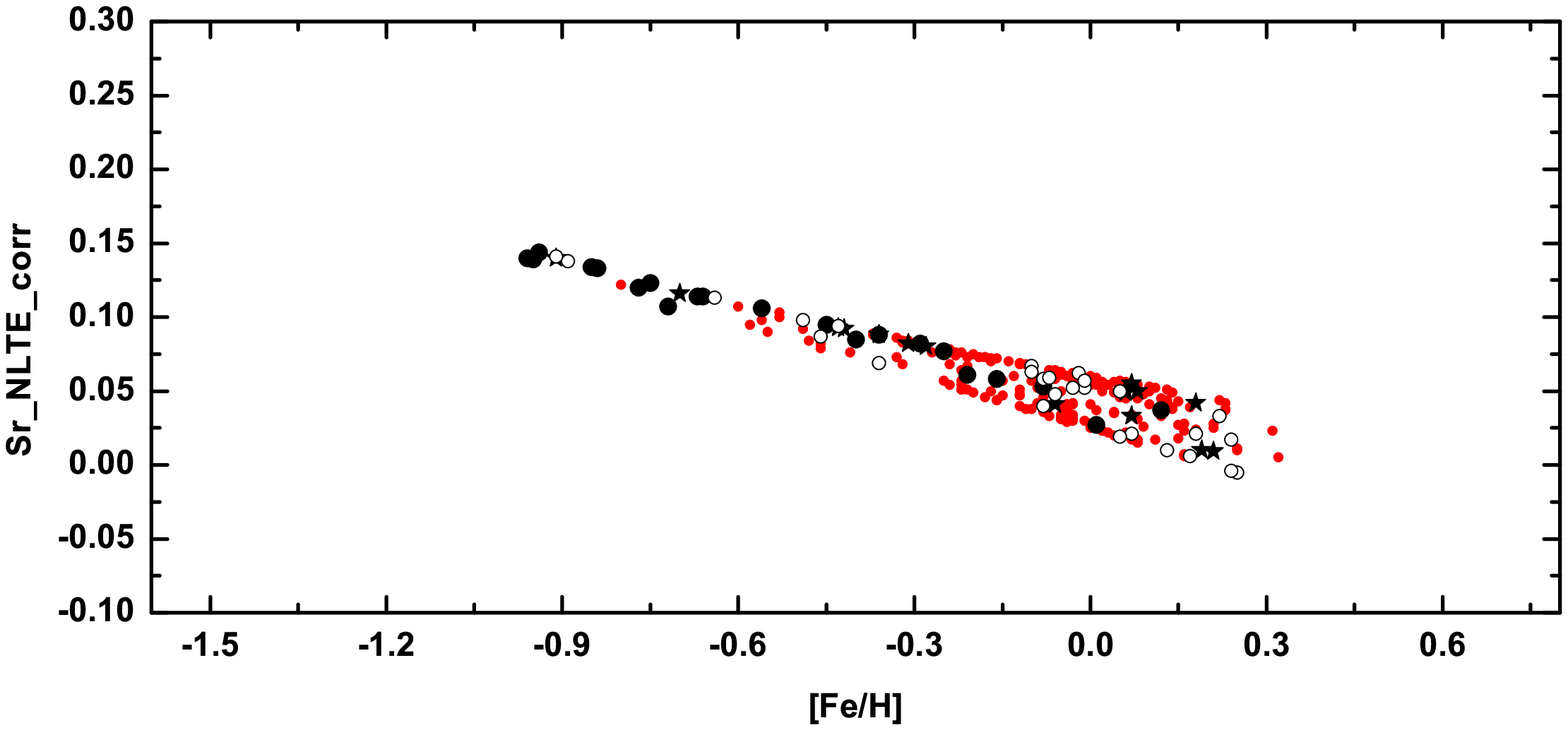}\\ 
\end{tabular}
\caption[]{Our determination LTE (upper panel) Sr abundances (thin disc - red circles, thick disc - black circles, Hercules stream -asterisks, unclassified - open circles) and a comparison with models of \cite{bisterzo:14} (thin disc - blue line, thick disc - doted blue line) and \cite{travaglio:04} (thin disc - black line, thick - black dashed line, only $s$-process contribution - black dotted line). NLTE corrections are from~\cite{bergemann:12} (bottom panel).}
\label{sr_LTE_NLTE}
\end{figure*}

As discussed in the introduction, a zoo of different nucleosynthesis processes can contribute to the production of Sr stable isotopes. Since spectroscopic observations can only determine element abundances for Sr, there exists no constraints on the isotopic pattern, excepting for the Sun. 
Therefore, it becomes difficult to disentangle the contribution from these different processes in stars in Galactic Archaeology studies \citep[e.g.,][]{yong:13}, where few or even only one single nucleosynthesis event could dominate the isotopic abundance pattern. 
Sr is often used as a tracer of the LEPP enrichment in metal-poor stars \citep[e.g.,][]{montes:07}. However, Sr elemental observations may lead to different interpretations.
\cite{froehlich:06} suggested the $\nu$p-process as source of nuclei up to A=90 or slightly beyond, originated in the neutrino-driven winds from forming neutron stars in CCSNe \citep[see also more recent investigations by][]{martinez:14, eichler:17}.
Due to the electron fraction $Y_e$ larger than 0.5, the $\nu$p-process acts on the proton-rich side of the valley of stability, producing a non-solar isotopic pattern.
The weak $s$-process, activated by the $^{22}$Ne neutron source in massive stars, is usually metallicity-dependent and negligible at low metallicities. 
However, in fast-rotating massive stars $^{14}$N can be made by rotational mixing, leading to a primary production of $^{22}$Ne in He-burning.
The $s$-process production in fast-rotating massive stars has been investigated \citep[][]{pignatari:08,frischknecht:16, meynet:17,choplin:17,nishimura:17,prantzos:18}, and considered by GCE modeling \citep[][]{cescutti:15a,bisterzo:17,prantzos:18}. 
On the other hand, as mentioned in the introduction the existence of the LEPP component is controversial \citep[][]{cristallo:15,trippella:16,prantzos:18}. A consistent set of observations over a large sample of stars such as the one presented in this work becomes instrumental to shed more light into this debate.

Our results for the Sr abundance obtained within the LTE approximation and the NLTE corrections \citep{bergemann:12} for our Sr determinations are shown in Fig. \ref{sr_LTE_NLTE}, in comparison to GCE model results from \cite{travaglio:04,bisterzo:17}. 
This GCE model \citep{travaglio:04} follows the composition of stars, stellar remnants, interstellar matter (atomic and molecular gas), and their mutual interaction, in the three main zones of the Galaxy, halo, thick disc, and thin disc.  The chemical enrichment takes into account the $s$-process yields from AGB stars, the $r$ contribution from massive stars (estimated with the residual method Nr = N$_\odot$ −- Ns), and the primary LEPP contribution. 
As discussed in \citep{travaglio:04}, the impact of AGB uncertainties on GCE computations may be partially reduced by assuming a range of $^{13}$C-pocket strengths, according to the $s$-process spread observed in disc stars and in presolar meteoritic SiC grains. The $r$ contribution was assumed to derive from SNeII of 8 -10 M$_\odot$.
 Nevertheless we do not exclude different hypotheses to explore the chemical origin of the Galactic halo (e.g., see discussion in Section 1). The LEPP contribution was evoked to explain the missing abundance of solar Sr; the ﬂat [Sr/Fe] trend observed at low metallicities suggested that LEPP is a primary process, likely occurring in CCSNe with an extended range of mass progenitors compared to the main $r$-process. 
In Fig. \ref{sr_LTE_NLTE}, the Galactic disc predictions by \cite{travaglio:04} are represented by black line for thin disc and dashed black line for thick disc. Note that the models which consider only the contribution in neutron-capture enrichment from $s$- and $r$- processes do not reproduce the observations at low metallicity (black dotted line).

\begin{figure*}
\begin{tabular}{c}
\includegraphics[width=16cm]{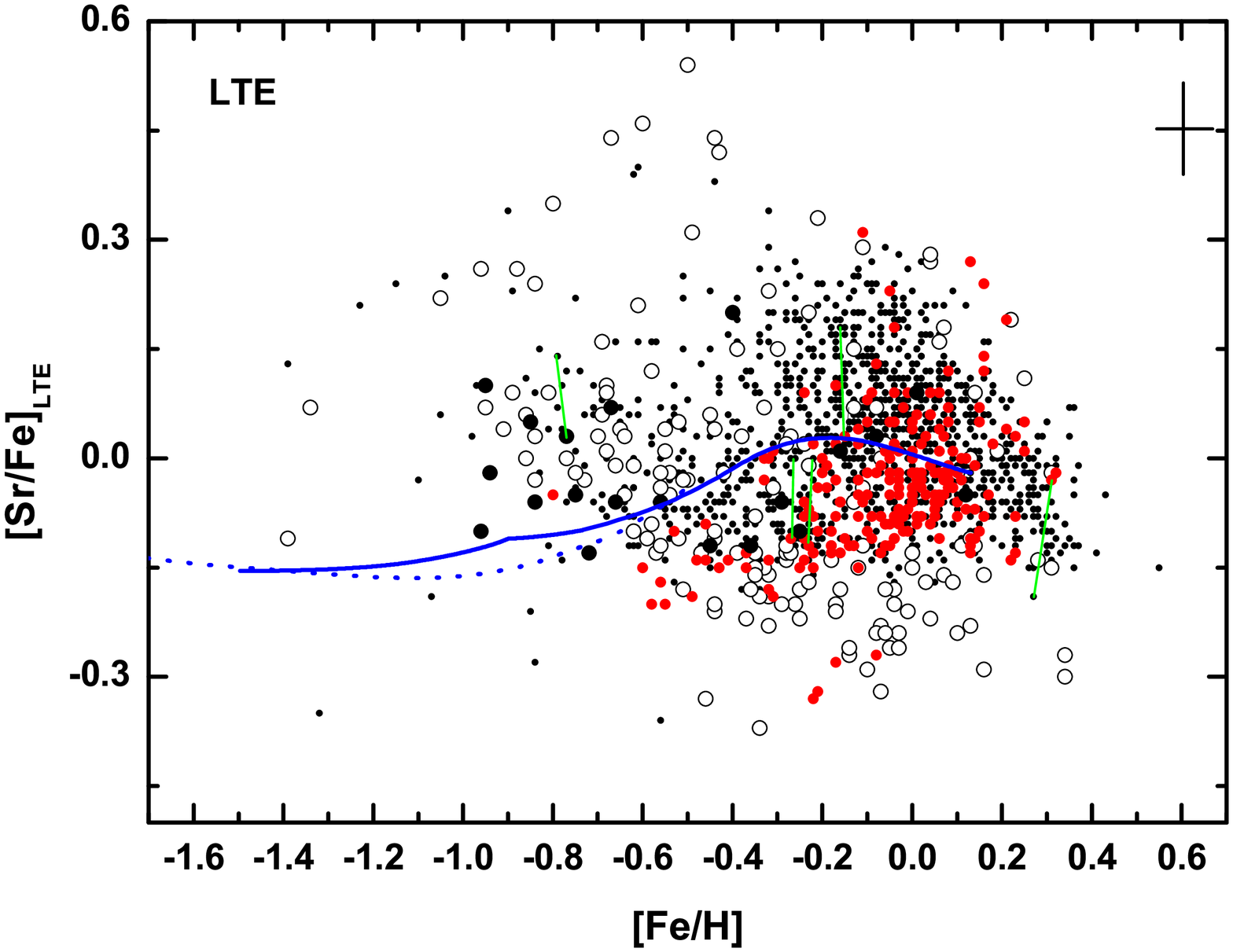}\\  
\end{tabular}
\caption[]{A comparison of our [Sr/Fe]LTE  (thin disc - red circles, thick disc - black circles)  with the data of \cite{battistini:16} (open circles), \cite{delgado:17} (points) and the chemical evolution prediction by \cite{bisterzo:17} (thin disc - blue line, thick disc - doted blue line).The values of the Sr abundance obtained  by us and in  other works that are different by more than 0.1 dex are connected (marked) by green lines.}
\label{comp_obs}
\end{figure*}

In Figs. \ref{sr_LTE_NLTE}, \ref{comp_obs}, GCE calculations by \cite{bisterzo:17} are compared with the observations (thin disc - blue line, thick disc - blue dotted line). 
\cite{bisterzo:17} simulations included new stellar yields and GCE parameters set compared to \cite{travaglio:04}.
In particular, GCE Fe predictions by \cite{bisterzo:17} are obtained by using SNIa stellar yields by \citep{travaglio:05}, coupled with an updated treatment of the delayed-time distribution function as suggested by \cite{kobayashi:98}, \cite{kobayashi:15}, \cite{greggio:05}, \cite{matteucci:09}, in which we assume a dominant SNIa contribution starting from [Fe/H] $>$ -- 1. 

\cite{bisterzo:17} also investigate the impact on GCE simulations of the internal structure of the $^{13}$C pocket, that is one of the major uncertainties for the $s$-process production in AGB stars.
Considering these uncertainties, the authors confirmed their earlier results \citep[][]{travaglio:04}, 
where an additional LEPP contribution is required in order to represent the solar $s$-process abundances of isotopes from A = 90 to 130 \citep[solar LEPP,][]{montes:07}.
\cite{bisterzo:17} also discussed the impact of the $s$-process yields from fast-rotating massive stars, with a contribution up to $\sim$ 17$\%$ to solar Sr 
\citep[$s$-process yields from fast-rotating massive stars yields by][]{frischknecht:16}.
Therefore, according to those calculations, the maximum $s$-process production of Sr is Sr$_s$ $\sim$ 90$\%$.
Instead, $s$-process isotopes and elements with 90 $<$ A $<$ 130 are marginally affected by this additional source of $s$-process, with variations within the solar uncertainties. 

GCE simulations presented in the figure can reproduce Sr production in the galaxy and the solar abundances of Sr. In particular, by considering also the contribution from fast-rotating massive stars the [Sr/Fe] abundance in thin-disc stars is better reproduced compared to \cite{travaglio:04}.
On the other hand, the production of Y and heavier LEPP elements is not obtained, possibly made by a combination of other nucleosynthesis processes. 

\begin{figure}
\begin{tabular}{c}
\includegraphics[width=8.4cm]{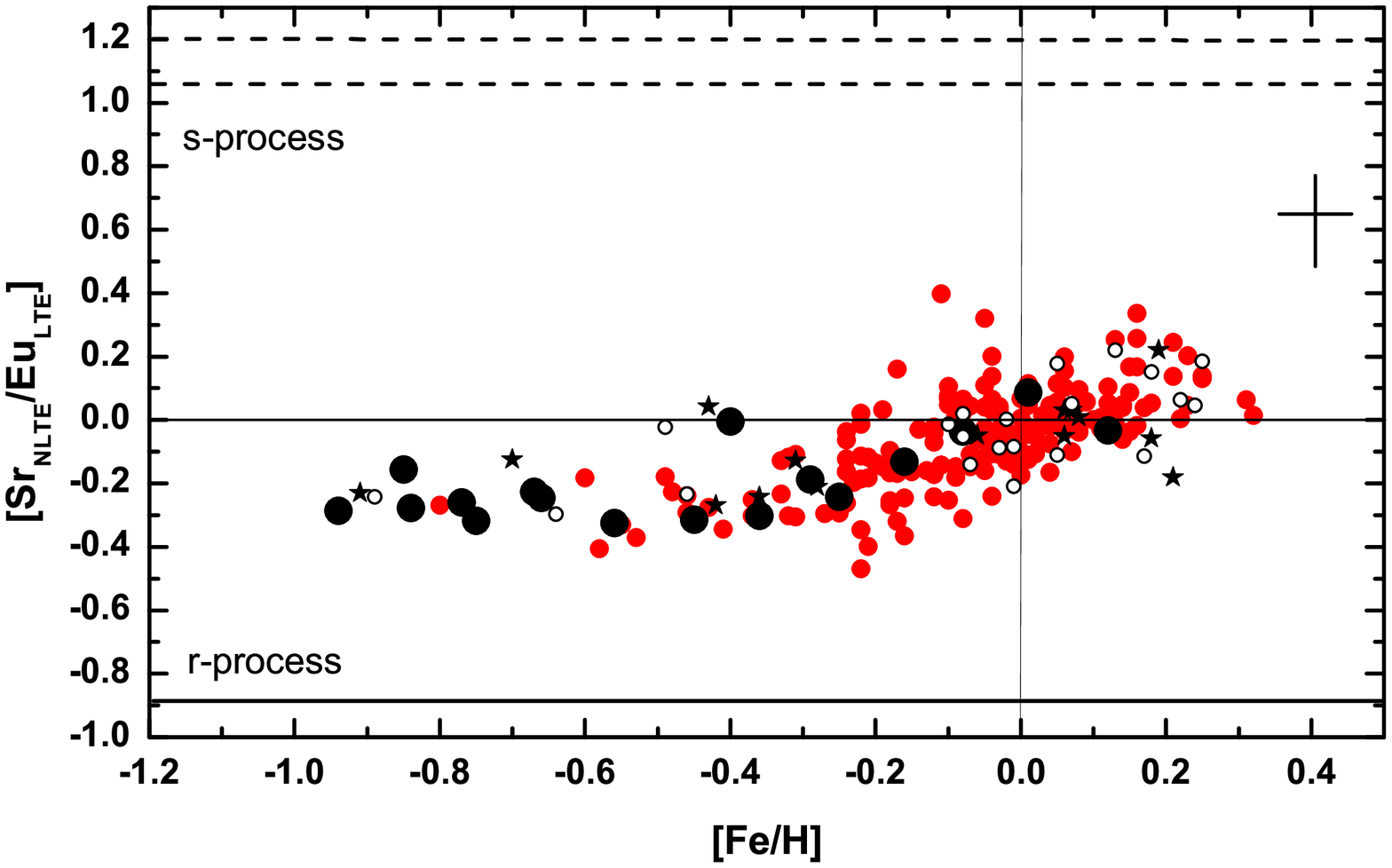}\\ 
\includegraphics[width=8.4cm]{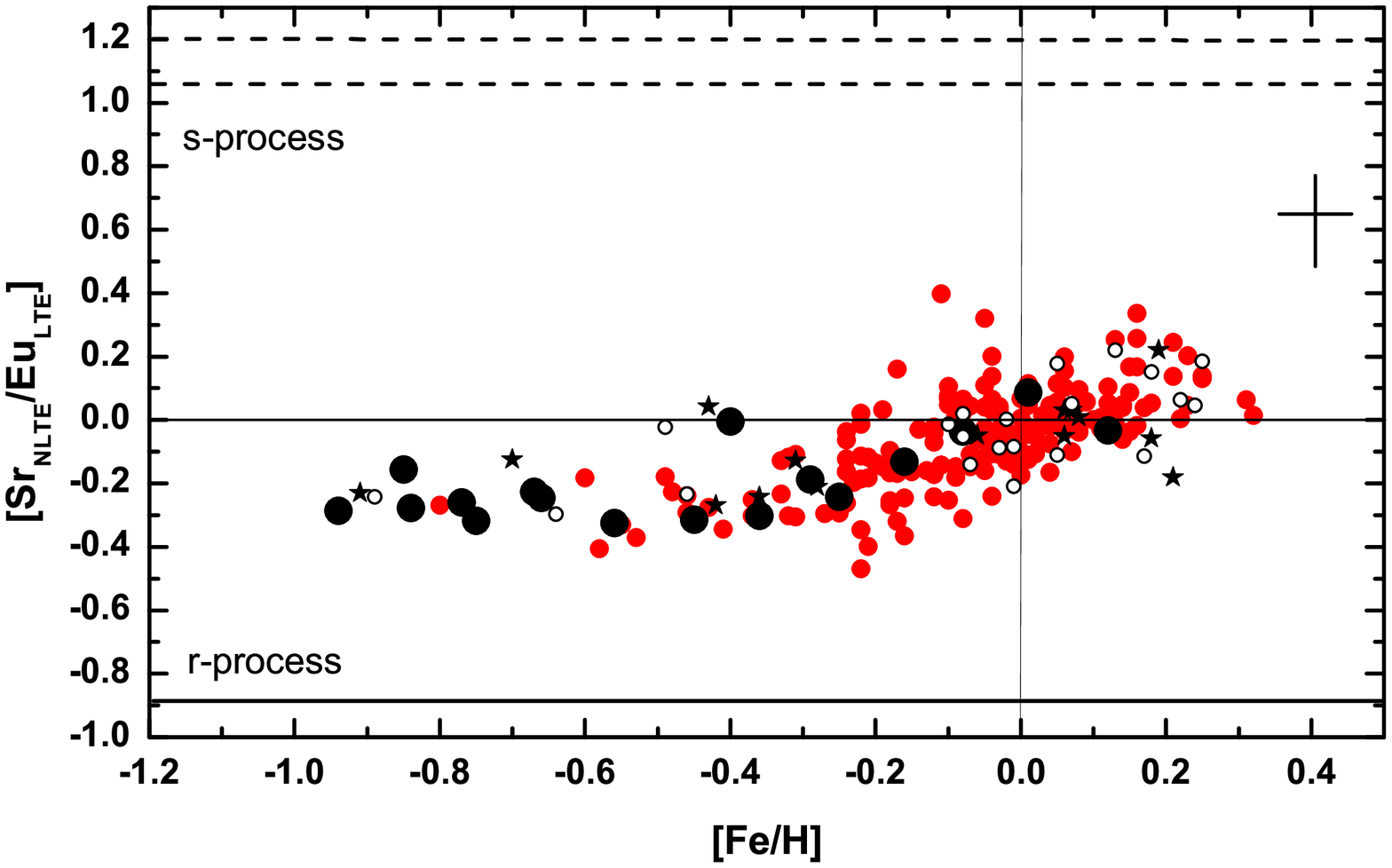}\\ 
\includegraphics[width=8.4cm]{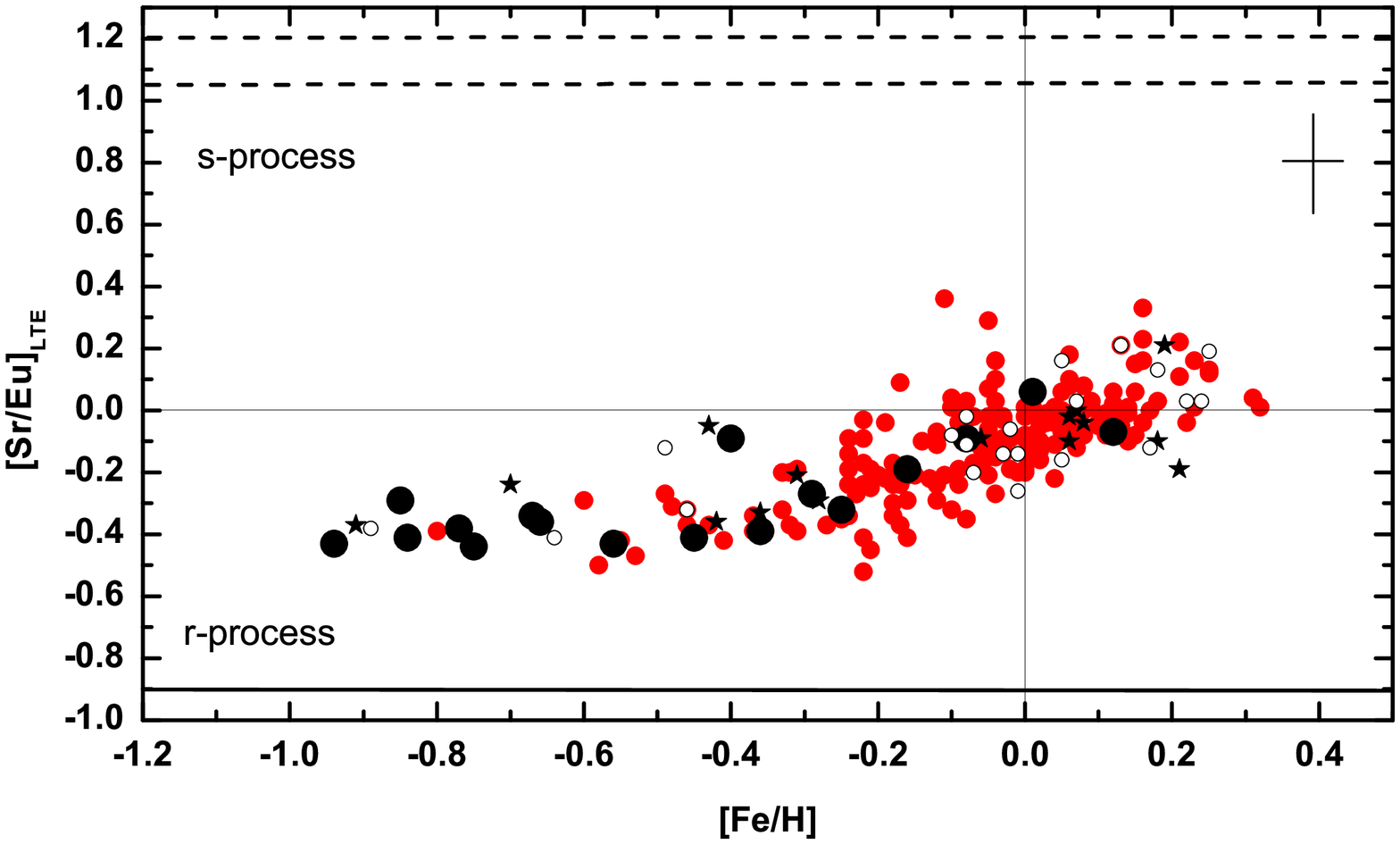}\\ 
\end{tabular}
\caption[]{Dependencies of [Sr/Ba] (NLTE),  [Sr/Eu] (Sr NLTE data) and [Sr/Eu] (Sr LTE data) vs.[Fe/H] with $r$-, $s$- process introduction from \cite{bisterzo:17}. The $s$-process signatures (pure AGB $s$-process production and including $s$-process contribution from massive stars) and $r$-process signatures are included: [Sr/Ba]$_{\rm r}$ = -0.10; [Sr/Eu]$_{\rm r}$ = -0.89; [Ba/Eu]$_{\rm r}$ = -0.80; [Sr/Ba]$_{\rm s}$ = -0.09 (pure AGB)- +0.05 (AGB + massive stars); [Sr/Eu]$_{\rm s}$ = 1.06 (pure AGB) - 1.20(AGB + massive stars); [Ba/Eu]$_{\rm s}$ = 1.15. Notations are: thin disc stars marked  as red  circles, thick disc stars -- as black circles, Hercules stream stars -- as asterisks, non-classified stars -- as small open circles; $r$-process as solid line, $s$-process as dotted lines.}
\label{sr_ba_eu_fe}
\end{figure}

In Fig. \ref{sr_ba_eu_fe} we show the evolution of the [Sr/Ba] and [Sr/Eu] ratios with respect to [Fe/H]. The average observational error is reported in the figures. The abundances for Ba and Eu were taken from  \cite{mishenina:13}. Ba abundances were computed under the NLTE approximation in our earlier studies  
\citep[][]{korotin:11,mishenina:13}. Ba abundances in dwarf stars were determined from Ba II 4554, 5853, 6141 and 6496 \AA~  while the LTE Eu abundance was derived from the line at 6645 \AA~ \citep{mishenina:13}. The NLTE profiles of the Ba lines were computed using a modified version of the MULTI code \citep{carlsson:86}; all modifications have been described in detail in \citep{korotin:99}. The Ba model in this study contains 31 levels of Ba~{\sc I}, 101 levels of Ba~{\sc II} with n $<$ 50 and the ground level of the Ba~{\sc III}. The analysis covers 91 bound-bound transitions. The NLTE Ba calculations have been described in detail in \citep{korotin:11}. In order to verify the effect of the LTE deviations on the Sr abundance, as well as on their relationship with other elemental abundances, we have plotted [Sr/Eu] vs [Fe/H] using both the NLTE and LTE Sr abundances. As can be seen in Fig. \ref{sr_ba_eu_fe}, there is no significant difference. 
A pure $r$-process signature has been indicated for both [Sr/Ba] and [Sr/Eu] assuming that Sr$_{\rm r}$ = 9$\%$ of the solar Sr abundance. This estimation is based on observations of very metal-poor $r$-process rich stars \citep{travaglio:04,mashonkina:14,roederer:14a}. We use then Ba$_{\rm r}$ = 15$\%$ of the solar Ba content and Eu$_{\rm r}$ = 94$\%$ of the solar Eu abundance \citep[using the residual method and the GCE $s$-process calculations by][]{bisterzo:17}.
The $r$-process contribution to Ba and Eu in the solar composition are derived 
by subtracting the $s$-process fractions from the solar abundances ($r$-process residuals method). 

\begin{figure}
\begin{tabular}{c}
\includegraphics[width=8.4cm]{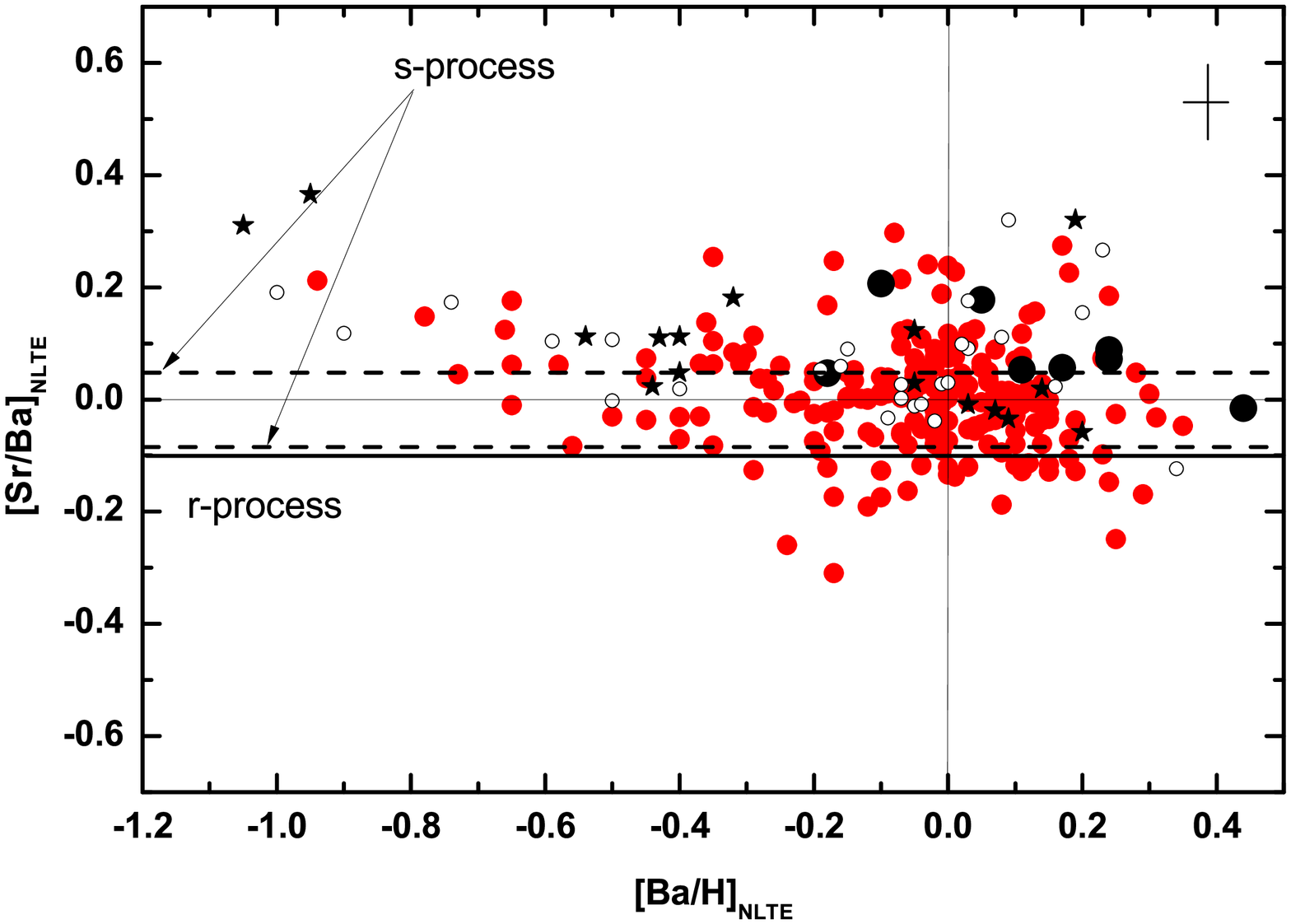}\\ 
\includegraphics[width=10cm]{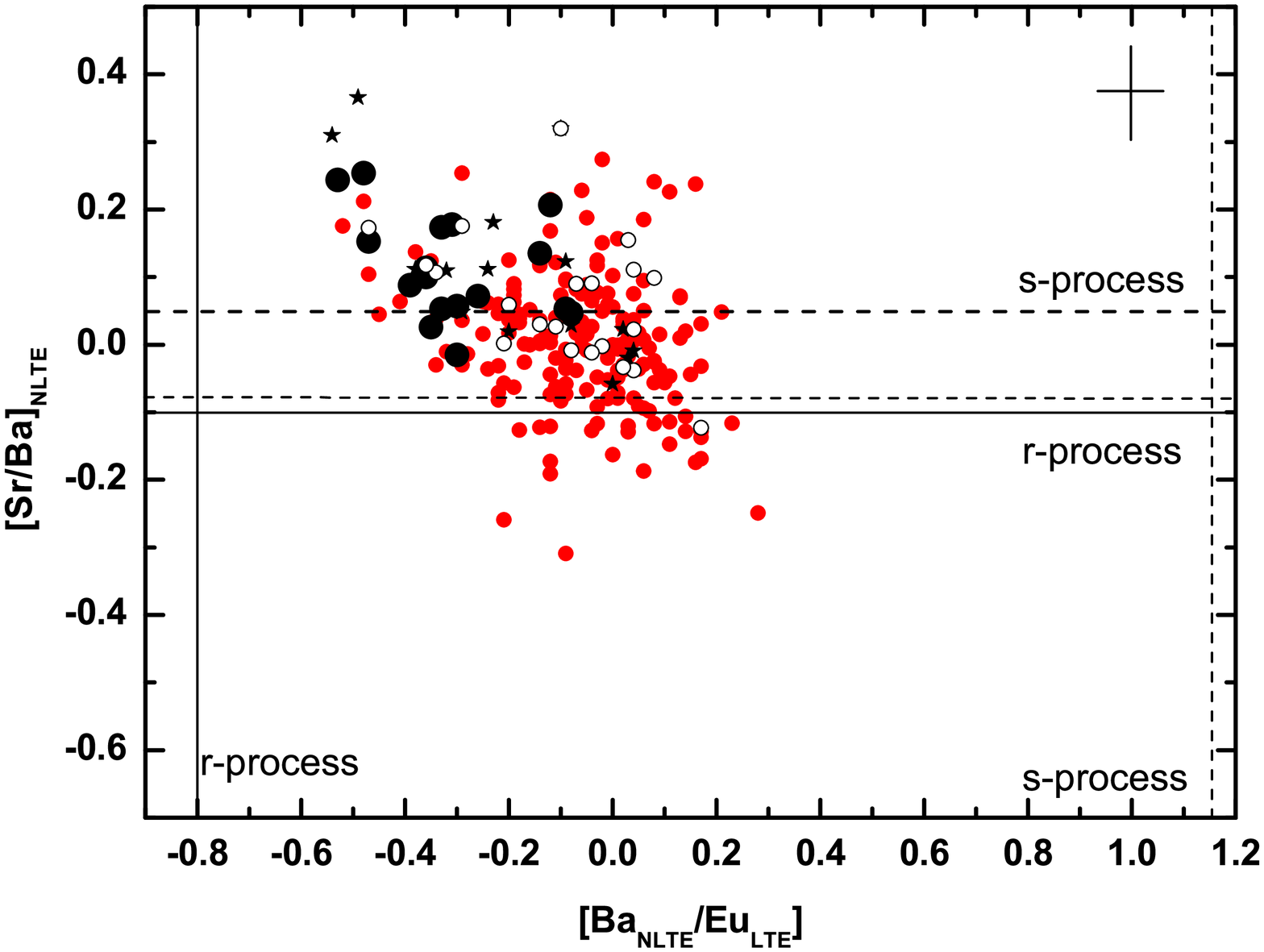}\\
\end{tabular}
\caption[]{The [Sr/Ba] ratio (NLTE Sr, Ba) is shown with respect to [Ba/H] (NLTE Ba) and [Ba/Eu] (NLTE Ba, LTE Eu with $r$-, $s$- process introduction from \cite{bisterzo:17}.  Notations are the same as in Fig. \ref{sr_ba_eu_fe}.}
\label{srba_baeu}
\end{figure}

The [Sr/Eu]$_{\rm r}$ ratio is well below any star observed in the Galactic disc, confirming that other early nucleosynthesis processes producing Sr are contributing. The [Sr/Ba]$_{\rm r}$ ratio is close to the solar ratio, and not much information can be derived. 
In Fig. \ref{srba_baeu}, the [Sr/Ba] (NLTE) ratio is shown with respect to the [Ba/Eu] (NLTE Ba, LTE Eu) ratio. 
Ratios consistent with the $r$-process production and the $s$-process contribution are shown for comparison. Most of the stars show abundance signatures consistent with a combined contribution of $s$-process and $r$-process. 
Also in the Galactic disc, we can see for a number of stars a possible signature similar to the stellar LEPP \citep[][]{montes:07}, where the [Sr/Ba] is larger than the $s$-process contributions and the $r$-process, and consistent with \cite{francois:07} results, where an anti-correlations of [Sr/Ba], [Y/Ba] and [Zr/Ba] ratios with −4.5 $<$ [Ba/H] $<$ −1.5 was obtained.
These results confirm the need of additional nucleosynthesis processes responsible for the synthesis of the first-peak elements.
\cite{andrievsky:11} have reanalyzed in NLTE approximation Sr and Ba abundances from \cite{francois:07} and have compared with the theoretical predictions of the LEPP model \citep{travaglio:04}. Their NLTE homogeneous determinations qualitatively confirm Sr, Y, Zr, Ba behaviour found in \cite{francois:07}, and enable one to robustly claim that the Sr abundances are generally higher than those predicted by the main $r$-process pattern. 
They have concluded that since the theoretical curve of a LEPP process is not far from the upper envelope of their data points, then an inhomogeneous mixing of the products of such a LEPP process with the products of the main $r$-process could explain the distribution of studied metal-poor stars. 
In our figures, the stars with highest [Sr/Ba] are less than 0.3 dex beyond the $s$-process prediction. This might be seen as a signature of the different nucleosynthesis processes contributing to Sr and discussed before for Galactic Archaeology studies, but this scatter is close to the [Sr/Ba] observational error. 
Four stars have a ratio of [Sr/Ba] (Sr and Ba abundances presented in NLTE approach) higher than 0.3 dex, namely HD64606 ([Sr/Fe] = 0.17, [Sr/Ba] = 0.31), HD139323 ([Sr/Fe] = 0.32, [Sr/Ba] = 0.32), HD144579 ([Sr/Fe] =0.11, [Sr/Ba] = 0.37) (Hercules stream) and one HD32147 ([Sr/Fe] = 0.28, [Sr/Ba] = 0.32) belongs to unclassified stars.
These stars have a different kinematics from the stars of thick and thin discs and this could give an occasion to consider their special enrichment with Sr. However, only two of them show some excess of Sr, slightly exceeding the determination errors. 
Interestingly, there are five stars with the [Sr/Ba] ratio falling outside the range of errors 
from $s$-process or $r$-process: 
HD 26923 ([Sr/Ba]=-0.249 LTE, 0.03 NLTE), HD45088 (-0.309, -0.269), HD53927 (-0.259, -0.279), HD127506 (-0.191, -0.231), HD141272 (-0.187, -0.047). 
The [Sr/Ba] ratios for each star are given in parentheses wherein the first value corresponds to the LTE Sr abundance and the second value corresponds to the NLTE Sr content; for Ba the NLTE abundance estimates are used in both cases. For three stars, HD45088, 53927 and 127506, these deviations are the same for both the LTE and NLTE Sr abundance determinations. Again, we are quite close to the error range limit.  The same uncommon signature is observed in few metal poor stars \citep[e.g.,][]{roederer:10,frebel:10,hansen:18}, indicating the contribution from different $r$-process components or some additional nucleosynthesis component which is not taken into account in this analysis. 

\begin{figure*}
\begin{tabular}{c}
\includegraphics[width=16cm]{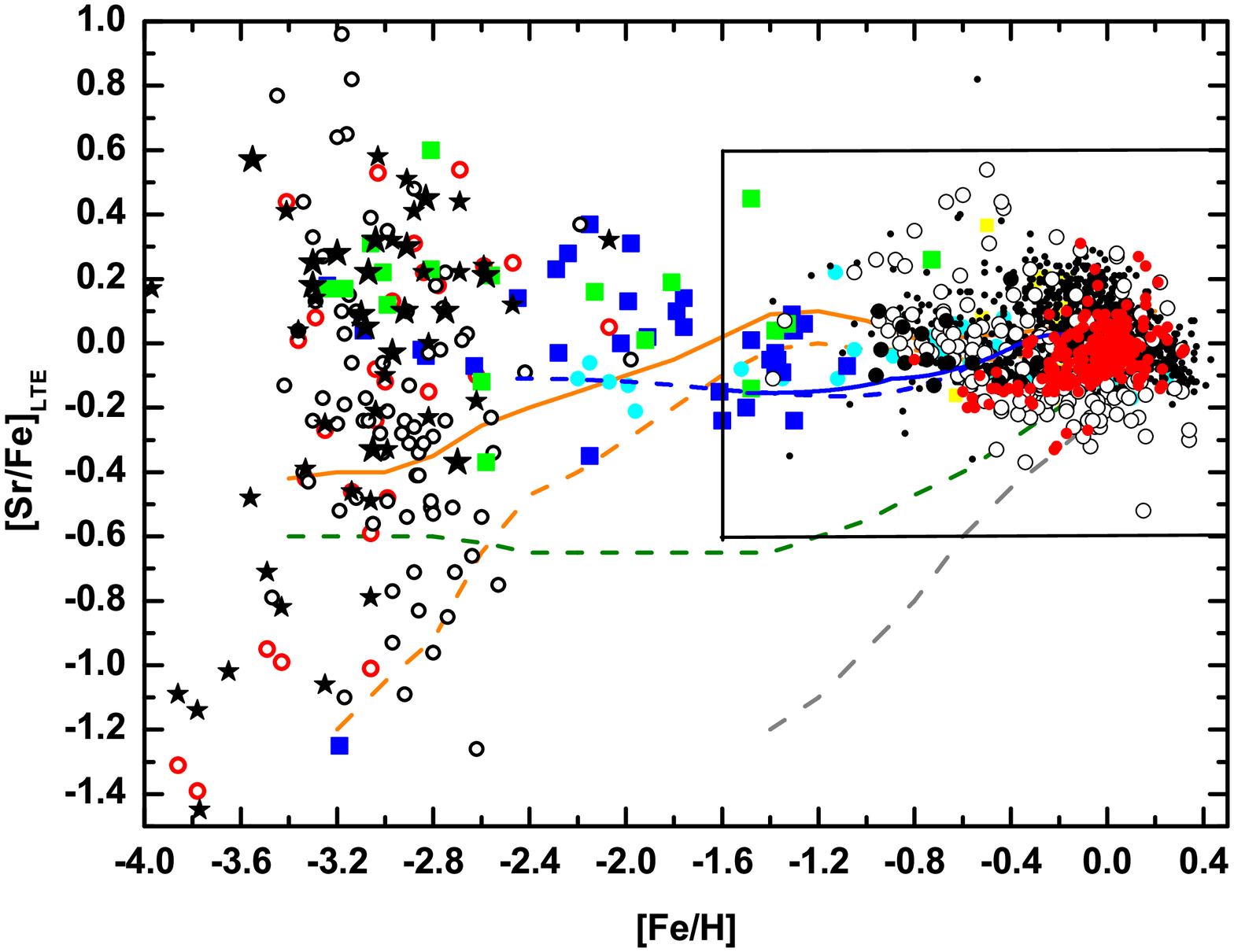}\\  
\includegraphics[width=16cm]{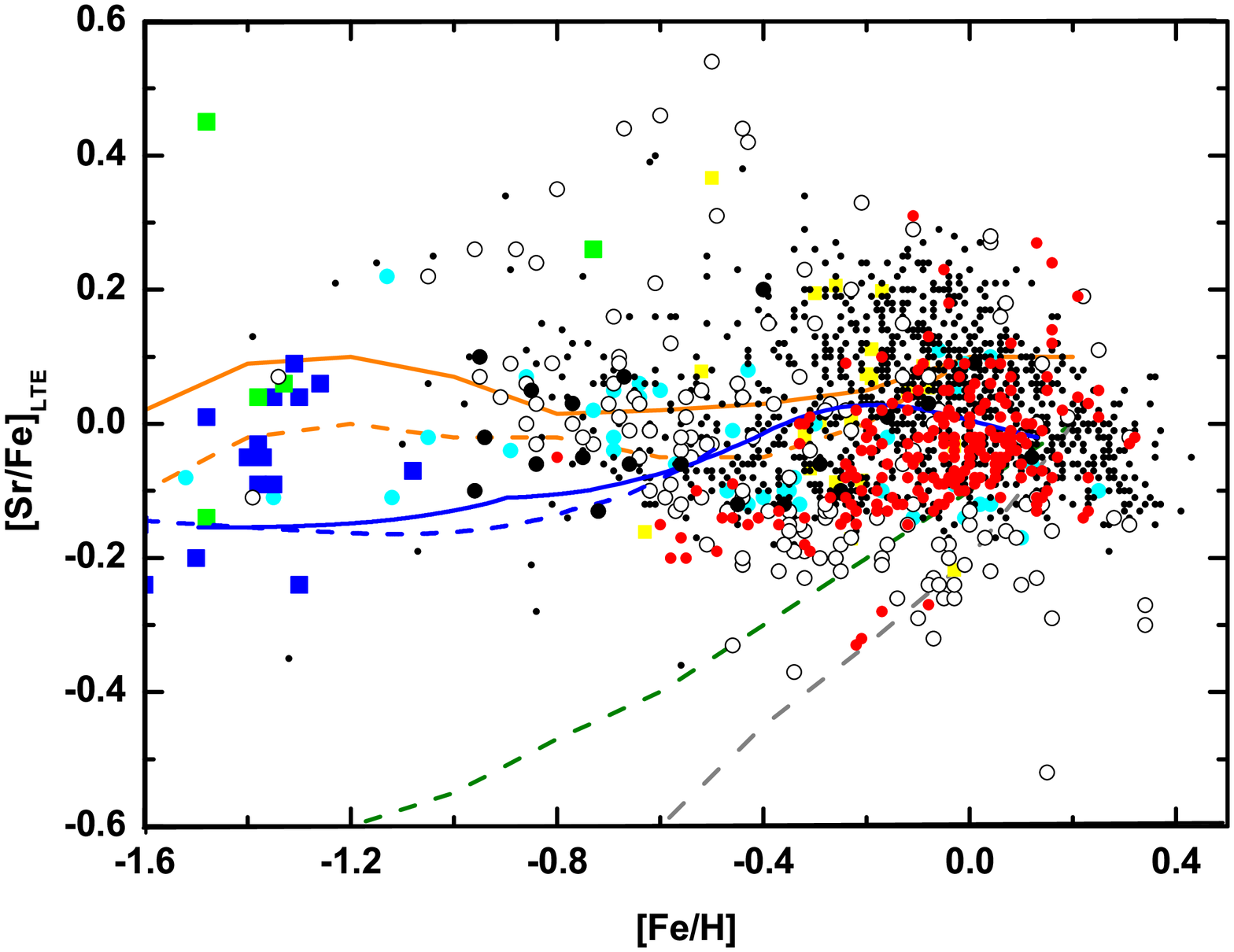}\\  
\end{tabular}
\caption[]{A comparison of our [Sr/Fe]LTE data and those of other works with GCE predictions with the chemical evolution prediction by \cite{bisterzo:17} and \cite{prantzos:18} (see model details in text).
The data from different literature sources are marked as follows: our thin disc data - as small red circles, our thick disc data - as black circles; data from \cite{mashonkina:01} - as cyan circles; data from \cite{brewer:06}  - as yellow circles, data from \cite{francois:07}  - as  big red open circles, data from \cite{andrievsky:11}  - as asterisks (small for turnoff stars, big for giants); \cite{ishigaki:13} - as blue squares; \cite{aoki:13}  - as open circles; \cite{hansen:13}  - as green squares;  \cite{battistini:16} -- as small open circles;  \cite{delgado:17} -- as small points. }
\label{comp_prant}
\end{figure*}

In Figure \ref{comp_prant} we have compared our [Sr/Fe] LTE data and those obtained in numerous studies within a large range of [Fe/H], with GCE predictions by \cite{bisterzo:17} and \cite{prantzos:18}.  
The computed evolution from  \cite{bisterzo:17} marked for thin disc as blue line and thick disc as doted blue line (Fig. \ref{comp_prant}).
The predicted evolution \citep{prantzos:18} is shown for the cases wherein different contributing sources were considered: i) low and intermediate mass (LIM) stars, rotating massive stars plus their fiduciary $r$-process (the baseline model, orange continuous curve); ii) LIM stars, non-rotating massive stars and $r$-process (green dashed curve); iii) LIM stars and non-rotating massive stars without $r$-process contribution (gray dashed curve); and iv) LIM stars plus rotating massive stars without the $r$-process contribution (orange dashed curve). The authors have drawn the conclusion that overall the computed [X/Fe] vs. [Fe/H] evolution for the $s$-elements is consistent with the evolution predictions made in the previous studies  \citep[e.g.][]{bisterzo:17} for metallicities typical of the disc ([Fe/H] $\geq$ --1.0), but the weak $s$-process in rotating massive stars plays a key role in the evolution of the $s$-elements at low metallicity. Note the extra source of neutron-capture elements required to explain the solar abundances which led  \cite{travaglio:04} to postulate as additional process (LEPP) could apparently be explained by \cite{prantzos:18} as a contribution from rotating massive stars.

\begin{figure*}
\begin{tabular}{c}
\includegraphics[width=16cm]{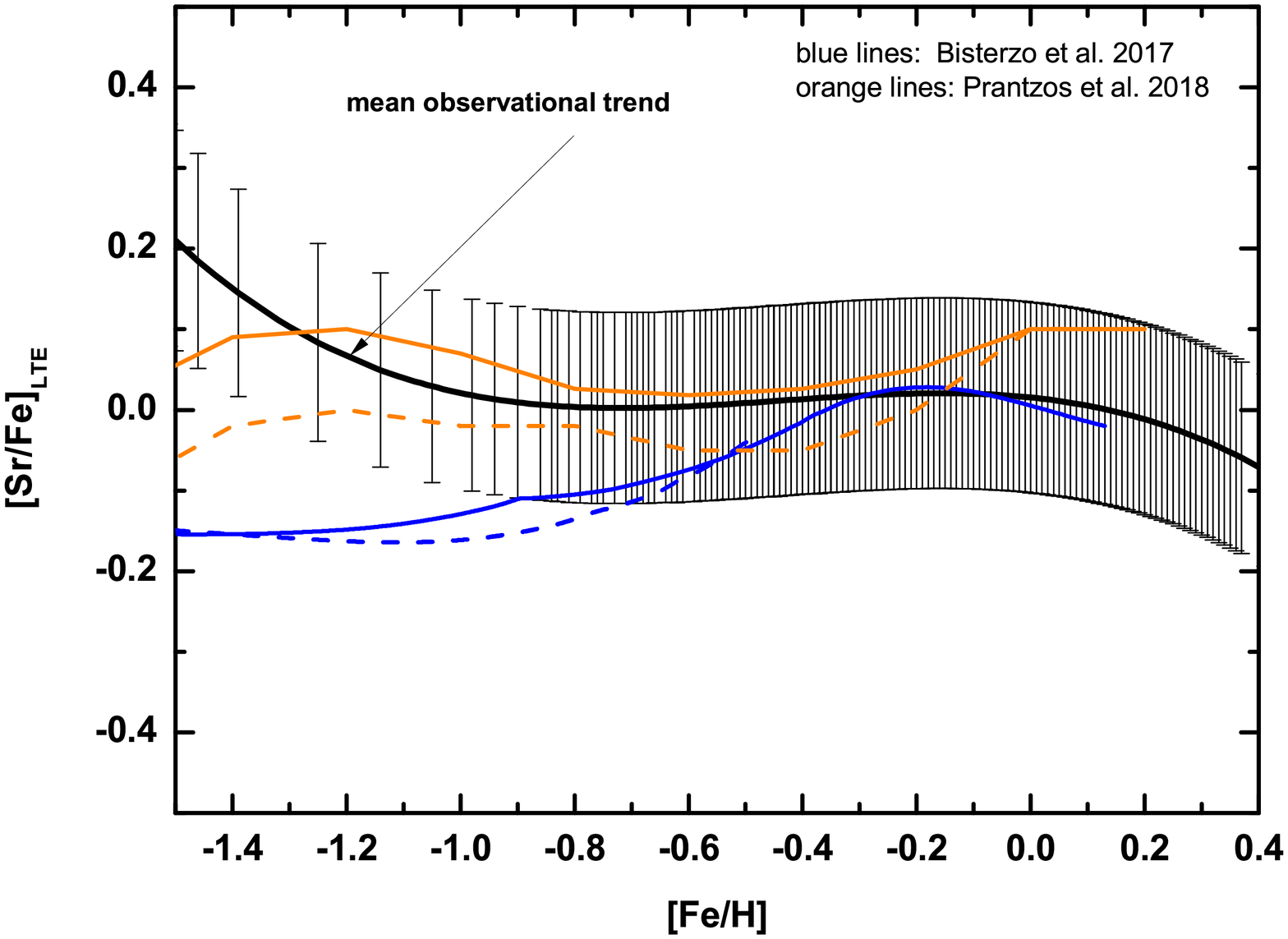}\\  
\end{tabular}
\caption[]{A comparison of the average observational [Sr/Fe]LTE data trend with the chemical evolution prediction by \cite{bisterzo:17} and \cite{prantzos:18} and in Galactic disc range of [Fe/H] . }
\label{trend1}
\end{figure*}

As can be seen in the figures, there is a large scatter of Sr abundances at all metallicities, including the near-solar ones, which are of specific interest in this study. The resulting spread exceeds the observation errors, as well as the differences obtained in various studies applying different approaches (e.g. using the LTE or NLTE assumptions). In order to evaluate the description of the observational data by various calculations (using different models) of GCE, we have presented our observations and those obtained by \cite{battistini:16} and \cite{delgado:17} as a single data set and expressed them as a third degree polynomial to plot versus the average observational trend.  Fig. \ref{trend1} illustrates the average observational trend with the error function determination by a polynomial, as well as models developed by \cite{bisterzo:17} and \cite{prantzos:18}. We have displayed the predicted evolution of \cite{bisterzo:17} (for thin disc as blue line and thick disc as doted blue line) and those of \cite{prantzos:18} for the cases wherein different contributing sources were considered: a) low and intermediate mass (LIM) stars, rotating massive stars and $r$-process contribution (the baseline model, orange continuous curve); b) LIM stars plus rotating massive stars without the $r$-process contribution (orange dashed curve). Indeed the computed [Sr/Fe] vs. [Fe/H] evolution is well consistent with the predictions made by  \cite{bisterzo:17} and \cite{prantzos:18} for metallicities typical for the disc stars ([Fe/H] $\geq$ --1.0). The main difference between the adopted models is due to the different contribution from massive rotating stars to the chemical enrichment. However, differences in the Sr abundance evolution in these two models are still within the accuracy of observations.

\section{Conclusions}
\label{sec: conclusions}

We present a new set of the Sr abundances measured for 276 stars, including 212 thin disc stars, 21 thick disc stars, 16 Hercules stream stars and 27 non-classified stars. By the time this study began, the Sr abundances had been determined for less than 2\% of the stars in our sample.  
The LTE approach was employed to estimate the abundances, whereby the departures from LTE were determined using the results of \cite{bergemann:12}; the average NLTE correction was 0.15 dex. Comparison of our data with those of other authors showed good agreement between them, with only five stars with a departure in the [Sr/Fe] by more than 0.05 dex. 

We obtain an observational scatter in the Sr abundance in the order of 0.2 -- 0.3 dex is measured, beyond observational errors (0.12 -- 0.17 dex, see Table \ref{errors}), 
in agreement with previous works. For thin disc stars we obtain a scatter -0.28 $\lesssim$ [Sr/Fe] $\lesssim$ 0.34, which is higher than the same range measured for more metal-poor thick-disc stars (from -0.03 $\lesssim$ [Sr/Fe] $\lesssim$ 0.26 dex). However, our sample of thick disc stars is too limited to draw robust conclusions. No significant trend is observed for the [Sr/Fe] evolution with respect to [Fe/H]. We note that there is
no significant difference between the LTE and NLTE trends as can be seen in Fig. \ref{sr_ba_eu_fe}.
 
We compared our results with the GCE calculations by \cite{travaglio:04}, \cite{bisterzo:17} and \cite{prantzos:18}.
A number of stellar sources contributed to the production of Sr in stars. We considered the $s$-process contribution from AGB stars, from massive stars and from fast-rotating massive stars. The Sr$_{s}$ ranges from 69$\%$ of the solar Sr due to AGB stars, up to $\sim$90$\%$ where also massive star contribution is taken into account. Based on observations of metal-poor $r$-process rich stars, the contribution to the solar Sr from the $r$-process is smaller than 10$\%$. The LEPP contribution was evoked to explain the missing abundance of solar Sr; the [Sr/Fe] trend observed at low metallicities suggested that LEPP is a primary process, likely occurring in CCSNe with an extended range of mass progenitors compared to the main $r$-process.

We have explored the Sr production together with Ba and Eu. We showed that while most of the stars can be explained within the $s$-process and $r$-process residual paradigm, there is a fraction of stars with [Sr/Ba] higher than the upper limit of Sr $s$-process contribution. While this feature is quite common in old stars formed in the early Galaxy, the observed departure from the $s$-process limit of [Sr/Fe] is much weaker in Galactic disc stars, within the observational errors ($\sim$ 0.2 dex). 
We obtain also a small fraction of stars with [Sr/Ba] lower by up to 0.2 dex than the $r$-process. At least for three stars, both LTE and NLTE Sr abundances display the values near -0.3 dex. Of course, taking into account the determination errors, this value is not so large, somewhere around 0.1 dex, which may be due to the dispersion of strontium and barium in the disc.

Using stellar data from our sample, from \cite{battistini:16} and \cite{delgado:17}  we have studied the production of Sr. Observations have been compared with GCE simulations by \cite{bisterzo:17} and \cite{prantzos:18}.
We confirm that
the $s$ -process contribution from the AGB stars, massive stars and fast-rotating massive stars is the main source of the Sr enrichment in the Galactic disc, possibly augmented by a CCSN contribution. The contribution of the fast-rotating massive stars becomes more significant with decreasing metallicity. 

A significant scatter in the [Sr/Fe] ratio is also seen in metal-poor stars, possibly indicating the contribution from additional $r$-process components in the early Galaxy (e.g. a main $r$-process contribution from rare events at low metallicities) and in the Galactic disc.

\section*{Acknowledgements}
We thank Elisa Delgado Mena for providing the data before the publishing. 
TM, TG, MP, FKT and SK thank for the support from the Swiss National Science 
Foundation, project SCOPES No. IZ73Z0$_{}$152485.
MP acknowledges significant support to NuGrid from NSF grants PHY 09-22648 
(Joint Institute for Nuclear Astrophysics, JINA), NSF grant PHY-1430152 
(JINA Center for the Evolution of the Elements) and EU MIRG-CT-2006-046520.
MP acknowledges the support from the "Lendület-2014" Programme of the Hungarian 
Academy of Sciences (Hungary), from SNF (Switzerland), from STFC (UK, through 
the University of Hull Consolidated Grant ST/R000840/1) and from 
VIPER HPC facility at the University of Hull (UK). 
FKT acknowledges support from the European Research Council (FP7) under ERC 
Advanced Grant Agreement 321263 FISH. SB thanks JINA (ND Fund 202476) for financial support.
TM thanks to O. Chepizhko for discussions. Authors are thankful to the anonymous referee 
for her/his very useful 
comments.

\bibliography{MN-18-1772-MJ.R2}


\appendix
\section{}

\onecolumn
\clearpage

\begin {longtable}{rccccccccc}
\label{t_abund}\\
\caption[]{Stellar parameters and abundances of some n-capture elements.The obtained (LTE)  Sr abundances, the NLTE corrections from \cite{bergemann:12}, the NLTE Ba and LTE Eu abundance, and stellar parameters \cite{mishenina:13}.}\\
\hline 
\hline
HD/BD    & \Teff, K & \logg  & [Fe/H] & \Vt & [Sr/Fe]LTE & corr$_{NLTE}$ & [Sr/Fe]NLTE & [Ba/Fe]NLTE & [Eu/Fe] \\ \hline
\endfirsthead
\caption{continued.}\\
\hline
HD/BD    & \Teff, K & \logg  & [Fe/H] & \Vt & [Sr/Fe]LTE& corr$_{NLTE}$& [Sr/Fe]NLTE & [Ba/Fe]NLTE & [Eu/Fe] \\ 
\hline
\endhead
\hline
\endfoot
\hline
thin    &      &        &       &  &       &        &        &        &       \\
\hline
166	&5514	&4.6	&0.16	&0.6  & 0.14  &     	0.12  &  	0.16   &      0.12    &         -0.09   \\
1562	&5828	&4	&-0.32	&1.2  & 0     &    	0.18  &  	0.08   &      0       &                 \\
1835	&5790	&4.5	&0.13	&1.1  & 0.27  &    	0.14  &  	0.31   &      0.04    &         0.06    \\
3651	&5277	&4.5	&0.15	&0.6  & 0.07  &    	0.11  &  	0.08   &      -0.14   &         -0.08   \\
4256	&5020	&4.3	&0.08	&1.1  & 0.12  &    	0.11  &  	0.13   &      -0.16   &                \\
4307	&5889	&4	&-0.18	&1.1  &-0.12  &    	0.17  &  	-0.04  &      0.08    &         0.12    \\
4614	&5965	&4.4	&-0.24	&1.1  &-0.06  &    	0.17  &  	0.01   &      0.02    &      0.08    \\
5294	&5779	&4.1	&-0.17	&1.3  & 0.1   &         0.17   & 	0.17    &      0.15    &     0.01    \\
6660	&4759	&4.6	&0.08	&1.4  & 0.05  &    	0.11  &  	0.06   &      -0.15   &         -0.03   \\
7590	&5962	&4.4	&-0.1	&1.4  & 0.08  &    	0.16  &  	0.14   &      0.11    &         0.07    \\
7924	&5165	&4.4	&-0.22	&1.1  &-0.13  &    	0.15  &  	-0.07  &      -0.05   &       0.04    \\
8648	&5790	&4.2	&0.12	&1.1  &-0.07  &    	0.14  &  	-0.02  &      -0.04   &        -0.13   \\
9407	&5666	&4.45	&0.05	&0.8  &-0.07  &    	0.14  &  	-0.02  &      -0.02   &         -0.03   \\
9826	&6074	&4	&0.1	&1.3  &-0.08  &    	0.15  &  	-0.02  &      -0.02   &                 \\
10086	&5696	&4.3	&0.13	&1.2  &-0.13  &    	0.14   & 	-0.09   &      -0.06   &         -0.08   \\
10307	&5881	&4.3	&0.02	&1.1  &-0.04  &    	0.15  &  	0.01   &      -0.02   &       0.12    \\
10476	&5242	&4.3	&-0.05	&1.1  & 0.01  &    	0.14   & 	0.05    &      0       &       -0.06   \\
10780	&5407	&4.3	&0.04	&0.9  & 0.06  &    	0.13  &  	0.09   &      0.09    &      0.05    \\
11007	&5980	&4	&-0.2	&1.1  &-0.02  &    	0.17  &  	0.05   &      0.05    &        0.19    \\
11373	&4783	&4.65	&0.08	&1    & 0.07  &    	0.11  &  	0.08   &      -0.04   &      -0.01   \\
12846	&5766	&4.5	&-0.24	&1.2  &-0.08  &    	0.17  &  	0.00  &      -0.04   &     0.16    \\
13507	&5714	&4.5	&-0.02	&1.1  & 0.07  &    	0.15  &  	0.12   &      0.11    &     0.16    \\
14374	&5449	&4.3	&-0.09	&1.1  & 0.09  &    	0.15  &  	0.14   &      0.02    &    0.13    \\
16160	&4829	&4.6	&-0.16	&1.1  &-0.13  &    	0.14  &  	-0.08  &      -0.19   &     0.28    \\
17674	&5909	&4	&-0.14	&1.1  &-0.12  &    	0.17   & 	-0.05   &      -0.03   &        -0.02   \\
17925	&5225	&4.3	&-0.04	&1.1  & 0.18  &    	0.13  &  	0.21   &      0.03    &     0.08    \\
18632	&5104	&4.4	&0.06	&1.4  & 0.04  &    	0.12  &  	0.06   &      -0.04   &       -0.04   \\
18803	&5665	&4.55	&0.14	&0.8  &-0.01  &    	0.13  &  	0.02   &      0       &       -0.02   \\
19019	&6063	&4	&-0.17	&1.1  & 0.1   &         0.17  &  	0.17   &      0.17    &          \\
19373	&5963	&4.2	&0.06	&1.1  &-0.06  &    	0.15  &  	0.00  &      -0.03   &        0.03    \\
20630	&5709	&4.5	&0.08	&1.1  &-0.01  &    	0.14  &  	0.03   &      0.07    &               \\
22049	&5084	&4.4	&-0.15	&1.1  & 0.03  &    	0.14  &  	0.07   &      0.15    &      0.24    \\
22484	&6037	&4.1	&-0.03	&1.1  &-0.07  &    	0.16  &  	-0.01  &      0.03    &      0.02    \\
22556	&6155	&4.2	&-0.17	&1.1  &-0.03  &    	0.17  &  	0.04   &      0.04    &      0.21    \\
24053	&5723	&4.4	&0.04	&1.1  & 0.06  &    	0.14  &  	0.10   &      0.11    &       0.1     \\
24238	&4996	&4.3	&-0.46	&1    &-0.14  &    	0.18  &  	-0.05  &      -0.12   &      0.18    \\
24496	&5536	&4.3	&-0.13	&1.5  &-0.12  &    	0.16   & 	-0.06   &      -0.12   &        0.1     \\
25665	&4967	&4.7	&0.01	&1.2  &-0.09  &    	0.12  &  	-0.06  &      -0.03   &        0.06    \\
25680	&5843	&4.5	&0.05	&1.1  &-0.02  &    	0.15  &  	0.03   &      0.05    &        0.02    \\
26923	&5920	&4.4	&-0.03	&1    &-0.03  &    	0.16  &  	0.03   &      0.28    &      0       \\
28005	&5980	&4.2	&0.23	&1.1  & 0.03  &    	0.14  &  	0.07   &      0       &        -0.13   \\
28447	&5639	&4	&-0.09	&1.1  &-0.11  &    	0.16  &  	-0.05  &      0.03    &        0.13    \\
29150	&5733	&4.3	&0	&1.1  &-0.05  &    	0.15  &  	0.00   &      -0.03   &        0.04    \\
29310	&5852	&4.2	&0.08	&1.4  &-0.02  &    	0.15   & 	0.03    &      0.02    &                 \\
29645	&6009	&4	&0.14	&1.3  &-0.11  &    	0.15  &  	-0.06  &      -0.07   &         -0.1    \\
30495	&5820	&4.4	&-0.05	&1.3  & 0.05  &    	0.16  &  	0.11   &      0.19    &     0.07    \\
33632	&6072	&4.3	&-0.24	&1.1  &-0.06  &    	0.17  &  	0.01   &      0.18    &     0.18    \\
34411	&5890	&4.2	&0.1	&1.1  &-0.06  &    	0.15   & 	-0.01   &      -0.01   &         -0.01   \\
37008	&5016	&4.4	&-0.41	&0.8  &-0.14  &    	0.17  &  	-0.06  &      -0.24   &     0.28    \\
37394	&5296	&4.5	&0.09	&1.1  & 0.01  &    	0.12  &  	0.03   &      0.06    &       -0.02   \\
38858	&5776	&4.3	&-0.23	&1.1  &-0.12  &    	0.17  &  	-0.04  &      0.03    &     0.15    \\
39587	&5955	&4.3	&-0.03	&1.5  &-0.05  &    	0.16  &  	0.01   &      0.14    &     -0.03   \\
40616	&5881	&4	&-0.22	&1.1  &-0.13  &    	0.17  &  	-0.05  &      0.12    &       -0.04   \\
41330	&5904	&4.1	&-0.18	&1.2  &-0.12  &    	0.17  &  	-0.05  &      0.01    &     0.22    \\
41593	&5312	&4.3	&-0.04	&1.1  & 0.09  &    	0.14  &  	0.13   &      0.1     &    -0.07   \\
42618	&5787	&4.5	&-0.07	&  1  &-0.08  &    	0.16  &  	-0.02  &      0.02    &     0.09    \\
42807	&5719	&4.4	&-0.03	&1.1  &-0.04  &    	0.15  &  	0.01   &      0.11    &     0.05    \\
43587	&5927	&4.1	&-0.11	&1.3  &-0.06  &    	0.16  &  	0.00   &      -0.04   &     0.15    \\
43856	&6143	&4.1	&-0.19	&1.1  &-0.04  &    	0.17  &  	0.03   &      0.15    &     0.18    \\
43947	&6001	&4.3	&-0.24	&1.1  &-0.14  &    	0.17  &  	-0.06  &      0.06    &     0.2     \\
45088	&4959	&4.3	&-0.21	&1.2  &-0.32  &    	0.15  &  	-0.27  &      0.04    &     0.13    \\
47752	&4613	&4.6	&-0.05	&0.2  &-0.05  &    	0.13  &  	-0.02  &      -0.02   &    0.1     \\
48682	&5989	&4.1	&0.05	&1.3  &-0.08  &   	0.15  &  	-0.02  &      -0.08   &       -0.08   \\
50281	&4712	&3.9	&-0.2	&1.6  & 0     &    	0.15  &  	0.05   &      0       &               \\
50692	&5911	&4.5	&-0.1	&0.9  &-0.1   &         0.16  &  	-0.03  &      0.03    &   0.22    \\
51419	&5746	&4.1	&-0.37	&1.1  &-0.13  &    	0.19  &  	-0.04  &      -0.08   &     0.26    \\
51866	&4934	&4.4	&0	&1    & 0     &         0.12  &  	0.02   &      -0.07   &     0.02    \\
53927	&4860	&4.64	&-0.22	&1.2  &-0.33  &    	0.15  &  	-0.28  &      -0.02   &     0.19    \\
54371	&5670	&4.2	&0.06	&1.2  & 0.01  &    	0.14  &  	0.05   &      -0.01   &       0.03    \\
55575	&5949	&4.3	&-0.31	&1.1  &-0.19  &    	0.18  &  	-0.10  &      0.02    &     0.2     \\
58595	&5707	&4.3	&-0.31	&1.2  & 0.01  &    	0.18  &  	0.09   &      0.01    &     0.2     \\
59747	&5126	&4.4	&-0.04	&1.1  & 0.05  &    	0.13  &  	0.08   &      0.09    &     0.02    \\
61606	&4956	&4.4	&-0.12	&1.3  & 0.02  &    	0.14   & 	0.06    &      0.02    &       0.13    \\
62613	&5541	&4.4	&-0.1	&1.1  &-0.05  &    	0.15  &  	0.00   &      0       &       -0.06   \\
63433	&5693	&4.35	&-0.06	&1.9  &-0.09  &    	0.16  &  	-0.03  &      0.02    &      0.03    \\
64468	&5014	&4.2	&0	&1.2  & 0.05  &    	0.12  &  	0.07   &      -0.17   &                 \\
64815	&5864	&4	&-0.33	&1.1  & 0     &   	0.18  &  	0.08   &      0.07    &        0.32    \\
65874	&5936	&4	&0.05	&1.3  &-0.05  &    	0.15  &  	0.00   &      -0.07   &         -0.11   \\
68638	&5430	&4.4	&-0.24	&1.1  &-0.11  &       	0.17  &  	-0.04  &        0.05  &      0.08  \\   
70923	&5986	&4.2	&0.06	&1.1  &-0.02  &    	0.15  &  	0.03   &        -0.06 &            -0.12 \\ 
71148	&5850	&4.2	&0	&1.1  &-0.05  &    	0.15  &  	0.00   &        -0.01 &             -0.06 \\ 
72760	&5349	&4.1	&0.01	&1.1  & 0.06  &    	0.13  &  	0.09   &        0.04  &           0.05  \\ 
72905	&5884	&4.4	&-0.07	&1.5  &-0.01  &    	0.16  &  	0.05   &        0.11  &         0.01  \\ 
73344	&6060	&4.1	&0.08	&1.1  &-0.04  &    	0.15  &  	0.01   &        -0.02 &          -0.04 \\ 
73667	&4884	&4.4	&-0.58	&0.9  &-0.2   &   	0.19  &  	-0.10  &        -0.15 &         0.3   \\ 
75732	&5373	&4.3	&0.25	&1.1  & 0.01  &    	0.11  &  	0.02   &        -0.13 &          -0.11 \\ 
75767	&5823	&4.2	&-0.01	&0.9  &-0.07  &    	0.15  &  	-0.01  &        0.04  &              \\ 
76151	&5776	&4.4	&0.05	&1.1  &-0.06  &    	0.15   & 	-0.01   &        -0.03 &            -0.06 \\ 
79969	&4825	&4.4	&-0.05	&1    &-0.02  &    	0.13  &  	0.01   &              &        0.07  \\ 
82106	&4827	&4.1	&-0.11	&1.1  & 0.31  &    	0.13  &  	0.34   &        0.11  &        -0.05 \\ 
82443	&5334	&4.4	&-0.03	&1.3  & 0.01  &    	0.14  &  	0.05   &        0.13  &        0.12  \\ 
87883	&5015	&4.4	&0	&1.1  &-0.06  &    	0.12  &  	-0.03  &        -0.05 &             0.02  \\ 
88072	&5778	&4.3	&0	&1.1  &-0.04  &    	0.15  &  	0.01   &        -0.03 &             0.15  \\ 
89251	&5886	&4	&-0.12	&1.1  &-0.08  &    	0.16  &  	-0.01  &        0.05  &           0.16  \\ 
89269	&5674	&4.4	&-0.23	&1.1  &-0.07  &    	0.17  &  	0.00   &        0.03  &         0.2   \\ 
91347	&5931	&4.4	&-0.43	&1.1  &-0.15  &    	0.19  &  	-0.05  &        -0.02 &          0.22  \\ 
94765	&5077	&4.4	&-0.01	&1.1  & 0.09  &    	0.13   & 	0.12    &        0.07  &                  \\ 
95128	&5887	&4.3	&0.01	&1.1  &-0.08  &    	0.15  &  	-0.02  &        -0.05 &            0     \\ 
97334	&5869	&4.4	&0.06	&1.2  &-0.05  &    	0.15  &  	0.00   &        0.13  &            -0.01 \\ 
97658	&5136	&4.5	&-0.32	&1.2  &-0.18  &    	0.16  &  	-0.11  &        -0.03 &          0.19  \\ 
98630	&6060	&4	&0.22	&1.4  &-0.14  &    	0.14  &  	-0.09  &        -0.09 &             -0.1  \\ 
101177	&5932	&4.1	&-0.16	&1.1  &-0.06  &    	0.17  &  	0.01   &        0.01  &        0.15  \\ 
102870	&6055	&4	&0.13	&1.4  &-0.11  &    	0.15  &  	-0.06  &        -0.03 &            -0.09 \\ 
105631	&5416	&4.4	&0.16	&1.2  &-0.08  &    	0.12  &  	-0.05  &        -0.02 &          -0.04 \\ 
107705	&6040	&4.2	&0.06	&1.4  &-0.11  &    	0.15  &  	-0.05  &        0.06  &          -0.05 \\ 
108954	&6037	&4.4	&-0.12	&1.1  &-0.05  &    	0.17  &  	0.02   &        0.11  &        0.06  \\ 
109358	&5897	&4.2	&-0.18	&1.1  &-0.13  &    	0.17  &  	-0.06  &        -0.05 &         0.04  \\ 
110463	&4950	&4.5	&-0.05	&1.2  & 0     &    	0.13  &  	0.03   &        0.04  &        0.09  \\ 
110833	&5075	&4.3	&0	&1.1  & 0.04  &    	0.12  &  	0.06   &        -0.04 &                   \\ 
111395	&5648	&4.6	&0.1	&0.9  & 0.02  &    	0.14  &  	0.02   &        0.19  &           0.02  \\ 
112758	&5203	&4.2	&-0.56	&1.1  & 0.17  &    	0.19  &  	-0.07  &        -0.22 &              \\ 
114710	&5954	&4.3	&0.07	&1.1  & 0.05  &    	0.15  &  	0.00   &        0.11  &          -0.03 \\ 
115383	&6012	&4.3	&0.11	&1.1  & 0.03  &    	0.15  &  	0.02   &        0.12  &           0.05  \\ 
115675	&4745	&4.45	&0.02	&1    & 0.02  &    	0.12  &  	0.00   &        -0.07 &           0.03  \\ 
116443	&4976	&3.9	&-0.48	&1.1  & 0.14  &    	0.18  &  	-0.05  &        -0.18 &        0.17  \\ 
116956	&5386	&4.55	&0.08	&1.2  & 0.03  &    	0.13  &  	0.00   &        0.05  &           0.04  \\ 
117043	&5610	&4.5	&0.21	&0.4  & 0.04  &    	0.13  &  	0.07   &        0.1   &           -0.07 \\ 
119802	&4763	&4	&-0.05	&1.1  & 0.23  &    	0.13  &  	0.26   &        0.02  &           -0.06 \\ 
122064	&4937	&4.5	&0.07	&1.1  & 0.03  &    	0.11  &  	0.04   &        -0.07 &           0.07  \\ 
124642	&4722	&4.65	&0.02	&1.3  &-0.03  &    	0.12  &  	-0.01  &        -0.02 &           0.1   \\ 
125184	&5695	&4.3	&0.31	&0.7  &-0.03  &    	0.12  &  	-0.01  &        0.04  &           -0.07 \\ 
126053	&5728	&4.2	&-0.32	&1.1  &-0.14  &    	0.18  &  	-0.06  &        -0.13 &         0.06  \\ 
127506	&4542	&4.6	&-0.08	&1.2  &-0.27  &    	0.14  &  	-0.23  &        -0.04 &         0.08  \\ 
128311	&4960	&4.4	&0.03	&1.3  & 0.03  &    	0.12  &  	0.05   &        -0.03 &           0.04  \\ 
130307	&4990	&4.3	&-0.25	&1.4  &-0.15  &    	0.16  &  	-0.09  &        0.08  &         0.2   \\ 
130948	&5943	&4.4	&-0.05	&1.3  &-0.03  &    	0.16  &  	0.03   &        0.15  &         0.07  \\ 
131977	&4683	&3.7	&-0.24	&1.8  & 0.09  &    	0.15  &  	0.14   &        -0.11 &        0.18  \\ 
135599	&5257	&4.3	&-0.12	&1    & 0.04  &    	0.15  &  	0.09   &        0.1   &        0.11  \\ 
137107	&6037	&4.3	&0	     &1.1  & 0.01  &    	0.16   & 	0.07    &        0.09  &                \\ 
139777	&5771	&4.4	&0.01	&1.3  &-0.03  &    	0.15  &  	0.02   &        0.14  &           -0.09 \\ 
139813	&5408	&4.5	&0	&1.2  &-0.02  &    	0.14  &  	0.02   &        0.15  &             0.12  \\ 
140538	&5675	&4.5	&0.02	&0.9  & 0.02  &    	0.15   & 	0.07    &        0.06  &             0.12  \\ 
141004	&5884	&4.1	&-0.02	&1.1  &-0.08  &    	0.16   & 	-0.02   &        0     &           0.11  \\ 
141272	&5311	&4.4	&-0.06	&1.3  &-0.09  &    	0.14  &  	-0.04  &        0.14  &          0.08  \\ 
142267	&5856	&4.5	&-0.37	&1.1  &-0.15  &    	0.19  &  	-0.06  &        -0.03 &          0.19  \\ 
144287	&5414	&4.5	&-0.15	&1.1  &-0.11  &    	0.15  &  	-0.05  &        -0.03 &                  \\ 
145675	&5406	&4.5	&0.32	&1.1  &-0.02  &    	0.10  &  	-0.01  &        -0.09 &           -0.03 \\ 
146233	&5799	&4.4	&0.01	&1.1  &-0.02  &    	0.15  &  	0.03   &        0.01  &            0.08  \\ 
149661	&5294	&4.5	&-0.04	&1.1  & 0.02  &    	0.14   & 	0.06    &        0.01  &             0.03  \\ 
149806	&5352	&4.55	&0.25	&0.4  & 0.05  &    	0.11   & 	0.06    &        0.05  &            -0.08 \\ 
151541	&5368	&4.2	&-0.22	&1.3  &-0.15  &    	0.16  &  	-0.08  &        -0.15 &             0.26  \\ 
153525	&4810	&4.7	&-0.04	&1    &-0.11  &    	0.13  &  	-0.08  &        0.04  &             0.16  \\ 
154345	&5503	&4.3	&-0.21	&1.3  &-0.1   &   	0.17  &  	-0.03  &        -0.05 &             0.15  \\ 
156668	&4850	&4.2	&-0.07	&1.2  &-0.13  &    	0.13  &  	-0.09  &        -0.13 &             0.05  \\ 
156985	&4790	&4.6	&-0.18	&1    &-0.1   &    	0.14  &  	-0.05  &        -0.09 &             0.2   \\ 
158633	&5290	&4.2	&-0.49	&1.3  &-0.19  &    	0.19  &  	-0.09  &        -0.16 &             0.08  \\ 
160346	&4983	&4.3	&-0.1	&1.1  & 0.05  &    	0.14  &  	0.09   &        -0.08 &             0.04  \\ 
161098	&5617	&4.3	&-0.27	&1.1  &-0.11  &    	0.17  &  	-0.03  &        -0.02 &           0.26  \\ 
164922	&5392	&4.3	&0.04	&1.1  &-0.01  &    	0.13  &  	0.02   &        -0.1  &             0.1   \\ 
165173	&5505	&4.3	&-0.05	&1.1  &-0.12  &    	0.15   & 	-0.07   &        -0.07 &             0.09  \\ 
165341	&5314	&4.3	&-0.08	&1.1  &-0.02  &    	0.14  &  	0.02   &        0.03  &           0     \\ 
165476	&5845	&4.1	&-0.06	&1.1  &-0.12  &    	0.16  &  	-0.05  &        -0.06 &                 \\ 
165670	&6178	&4	&-0.1	&1.5  &-0.1   &   	0.16  &  	-0.03  &        0.1   &                   \\ 
165908	&5925	&4.1	&-0.6	&1.1  &-0.15  &    	0.20  &  	-0.04  &        0.04  &            0.14  \\ 
166620	&5035	&4	&-0.22	&1    &-0.08  &    	0.15  &  	-0.02  &        -0.09 &            0.16  \\ 
171314	&4608	&4.65	&0.07	&1    &-0.02  &    	0.12   & 	 0.00     &        -0.09 &             0.1   \\ 
174080	&4764	&4.55	&0.04	&1    & 0.09  &    	0.12   & 	0.11    &        -0.01 &             0.13  \\ 
176377	&5901	&4.4	&-0.17	&1.3  &-0.08  &      	0.17  &  	-0.01  &        0.05  &       0.14  \\       
176841	&5841	&4.3	&0.23	&1.1  &-0.08  &    	0.13  &  	-0.04  &        -0.12 &           -0.09 \\ 
178428	&5695	&4.4	&0.14	&1    &-0.07  &    	0.14  &  	-0.03  &        0.04  &           0.03  \\ 
180161	&5473	&4.5	&0.18	&1.1  & 0.02  &    	0.12  &  	0.04   &        0.07  &           -0.01 \\ 
182488	&5435	&4.4	&0.07	&1.1  &-0.03  &    	0.13  &  	0.00   &        -0.07 &           -0.03 \\ 
183341	&5911	&4.3	&-0.01	&1.3  &-0.1   &   	0.16  &  	-0.04  &        -0.08 &         0.1   \\ 
184385	&5536	&4.45	&0.12	&0.9  & 0     &   	0.13  &  	0.03   &        0.07  &           -0.02 \\ 
185144	&5271	&4.2	&-0.33	&1.1  &-0.03  &    	0.17  &  	0.04   &        -0.02 &         0.17  \\ 
185414	&5818	&4.3	&-0.04	&1.1  &-0.11  &    	0.16   & 	-0.05   &        0.07  &           0.04  \\ 
186408	&5803	&4.2	&0.09	&1.1  &-0.04  &    	0.15  &  	0.01   &        -0.03 &           -0.05 \\ 
186427	&5752	&4.2	&0.02	&1.1  &-0.08  &    	0.15  &  	-0.03  &        -0.07 &           0.02  \\ 
187897	&5887	&4.3	&0.08	&1.1  &-0.03  &    	0.15  &  	0.02   &        0.03  &                 \\ 
189087	&5341	&4.4	&-0.12	&1.1  &-0.03  &    	0.15  &  	0.02   &        0.1   &          0.06  \\ 
189733	&5076	&4.4	&-0.03	&1.5  &-0.09  &    	0.13  &  	-0.06  &        -0.11 &          0.05  \\ 
190007	&4724	&4.5	&0.16	&0.8  & 0.12  &    	0.10  &  	0.12   &        -0.03 &            -0.04 \\ 
190406	&5905	&4.3	&0.05	&1    &-0.06  &    	0.15  &  	-0.01  &        0.05  &           -0.03 \\ 
190470	&5130	&4.3	&0.11	&1    & 0     &   	0.11  &  	0.01   &        -0.08 &            0.01  \\ 
190771	&5766	&4.3	&0.13	&1.5  &-0.12  &    	0.14  &  	-0.08  &        -0.07 &            -0.06 \\ 
191533	&6167	&3.8	&-0.1	&1.5  &-0.02  &    	0.16  &  	0.04   &        0.09  &            -0.06 \\ 
191785	&5205	&4.2	&-0.12	&1.2  &-0.15  &    	0.14  &  	-0.10  &        -0.24 &          0.14  \\ 
195005	&6075	&4.2	&-0.06	&1.3  & 0.01  &    	0.16  &  	0.07   &        0.06  &                \\ 
195104	&6103	&4.3	&-0.19	&1.1  &-0.01  &    	0.17  &  	0.06   &        0.2   &          0.03  \\ 
197076	&5821	&4.3	&-0.17	&1.2  & 0.02  &    	0.17  &  	0.09   &        0.08  &          0.21  \\ 
199960	&5878	&4.2	&0.23	&1.1  &-0.13  &    	0.14  &  	-0.09  &        -0.11 &                  \\ 
200560	&5039	&4.4	&0.06	&1.1  & 0.09  &    	0.12   & 	0.11    &        0.04  &            -0.09 \\ 
202108	&5712	&4.2	&-0.21	&1.1  &-0.04  &    	0.17  &  	0.03   &        0.1   &         0.15  \\ 
202575	&4667	&4.6	&-0.03	&0.5  &-0.02  &    	0.13   & 	0.01    &        0.09  &           0.1   \\ 
203235	&6071	&4.1	&0.05	&1.3  &-0.08  &    	0.16  &  	-0.02  &        -0.05 &           -0.01 \\ 
205702	&6020	&4.2	&0.01	&1.1  &-0.03  &    	0.16  &  	0.03   &        -0.03 &           -0.02 \\ 
206860	&5927	&4.6	&-0.07	&1.8  &-0.08  &    	0.16  &  	-0.01  &        0.05  &               \\ 
208038	&4982	&4.4	&-0.08	&1    & 0.13  &    	0.13  &  	0.16   &        0.09  &         0.1   \\ 
208313	&5055	&4.3	&-0.05	&1    &-0.09  &    	0.13  &  	-0.05  &        -0.09 &         -0.03 \\ 
208906	&5965	&4.2	&-0.8	&1.7  &-0.05  &    	0.22  &  	0.07   &        -0.14 &           0.34  \\ 
210667	&5461	&4.5	&0.15	&0.9  & 0.05  &    	0.12  &  	0.07   &        -0.04 &           -0.01 \\ 
210752	&6014	&4.6	&-0.53	&1.1  &-0.1   &    	0.2    &   	0.00      &        0.03  &         0.37  \\ 
211472	&5319	&4.4	&-0.04	&1.1  & 0.02  &    	0.14  &  	0.06   &        0.1   &               \\ 
214683	&4747	&4.6	&-0.46	&1.2  &-0.09  &    	0.18  &  	-0.01  &        0.06  &         0.28  \\ 
216259	&4833	&4.6	&-0.55	&0.5  &-0.2   &   	0.19   & 	-0.11   &        -0.1  &            0.22  \\ 
216520	&5119	&4.4	&-0.17	&1.4  &-0.28  &    	0.15   & 	-0.23   &        -0.2  &            0.09  \\ 
217014	&5763	&4.3	&0.17	&1.1  &-0.05  &    	0.14  &  	-0.01  &        -0.1  &            -0.05 \\ 
217813	&5845	&4.3	&0.03	&1.5  &-0.05  &    	0.15  &  	0.00   &        0.04  &            -0.01 \\ 
218868	&5547	&4.45	&0.21	&0.4  & 0.19  &    	0.12  &  	0.21   &        0.03  &            -0.03 \\ 
219538	&5078	&4.5	&-0.04	&1.1  &-0.08  &    	0.13  &  	-0.04  &        -0.06 &            0.06  \\ 
219623	&5949	&4.2	&0.04	&1.2  &-0.09  &    	0.15  &  	-0.03  &        0.01  &             0.13  \\ 
220182	&5364	&4.5	&-0.03	&1.2  &-0.02  &       	0.14  &  	0.02   &        0.07  &       0.1   \\           
220221	&4868	&4.5	&0.16	&0.5  & 0.24  &    	0.10  &  	0.24   &        0.02  &           -0.09 \\     
221851	&5184	&4.4	&-0.09	&1    &-0.08  &    	0.14  &  	-0.03 &        0.02  &          0.11  \\     
222143	&5823	&4.45	&0.15	&1.1  &-0.1   &   	0.14  &  	-0.05  &        0.09  &            -0.02 \\      
224465	&5745	&4.5	&0.08	&0.8  &-0.04  &    	0.14  &  	0.00   &        0.05  &            0.04  \\     
263175	&4734	&4.5	&-0.16	&0.5  &-0.06  &    	0.14  &  	-0.02  &        -0.13 &           0.23  \\     
BD12063	&4859	&4.4	&-0.22	&0.6  & 0.02  &    	0.15  &  	0.07   &        0.07  &           0.05  \\     
BD124499&4678	&4.7	&0	&0.5  & 0.04  &    	0.12  &  	0.06   &        0.02  &            0.24  \\ 

\hline  
thick disc   &  &  &  &  &  &  &  &  &    \\
\hline
245	&5400	&3.4	&-0.84	&0.7	&-0.06 & 0.23	&0.07	&0.02	& 0.35      \\       
3765	&5079	&4.3	&0.01	&1.1	& 0.09 & 0.12	&0.11	&-0.09	& 0.03      \\     
5351	&4378	&4.6	&-0.21	&0.5	& --   & 0.16	&--	&-0.33	&	0.08      \\     
6582	&5240	&4.3	&-0.94	&0.7	&-0.02 & 0.24	&0.12	&-0.12&	0.41      \\      
13783	&5350	&4.1	&-0.75	&1.1	&-0.05 & 0.22	&0.07	&-0.08	&	0.39      \\      
18757	&5741	&4.3	&-0.25	&1	&-0.1  & 0.17	&-0.02	&-0.08	&	0.22      \\                                                          
22879	&5972	&4.5	&-0.77	&1.1	& 0.03 & 0.22	&0.15	&0.05	&	0.41      \\       
65583	&5373	&4.6	&-0.67	&0.7	& 0.07 & 0.21	&0.18	&-0.07	&	0.41      \\       
76932	&5840	&4	&-0.95	&1	& 0.1  & 0.23	&0.23	&0.1    &   \\                                                                     
106516	&6165	&4.4	&-0.72	&1.1	&-0.13 & 0.20	&-0.02	&0.09	          \\      
 110897	&5925	&4.2	&-0.45	&1.1	&-0.12 & 0.19	&-0.02	&-0.01	&0.29      \\     
135204	&5413	&4	&-0.16	&1.1	& 0.01 & 0.15	&0.06	&-0.11	&0.2       \\      
152391	&5495	&4.3	&-0.08	&1.3	& 0.03 & 0.15	&0.08	&0.03	&0.12      \\      
157089	&5785	&4	&-0.56	&1	&-0.06 & 0.20	&0.04	&0.02	&0.37      \\     
157214	&5820	&4.5	&-0.29	&1	&-0.06 & 0.18	&0.02	&-0.05	&0.21      \\     
159062	&5414	&4.3	&-0.4	&1	& 0.2  & 0.18	&0.28	&0.15	&0.29      \\      
165401	&5877	&4.3	&-0.36	&1.1	&-0.12 & 0.18	&-0.02	&-0.12	&0.27      \\     
190360	&5606	&4.4	&0.12	&1.1	&-0.05 & 0.13	&-0.01	&-0.06	&0.02      \\     
201889	&5600	&4.1	&-0.85	&1.2	& 0.05 & 0.23	&0.18	&0.01	&0.34      \\      
201891	&5850	&4.4	&-0.96	&1	&-0.1  & 0.24	&0.04	&-0.06	&       \\     
204521	&5809	&4.6	&-0.66	&1.1	&-0.06 & 0.21	&0.05	&-0.06	&	0.3       \\
      
\hline 
 Hercules stream   &  &  &  &  &  &  &  &  &    \\
\hline  
13403	&5724	&4	&-0.31	&1.1	&-0.06	&0.18	&0.02	&-0.09	&0.15     \\
19308	&5844	&4.3	&0.08	&1.1	&-0.08	&0.15	&-0.03	&-0.01	& -0.04    \\
23050	&5929	&4.4	&-0.36	&1.1	&-0.08	&0.18	&0.00	&-0.04	&0.25     \\
30562	&5859	&4	&0.18	&1.1	&-0.08	&0.14	&-0.03	&0.02	&0.02     \\
64606	&5250	&4.2	&-0.91	&0.8	& 0.03	&0.24	&0.17	&-0.14	&0.4      \\
68017	&5651	&4.2	&-0.42	&1.1	&-0.1 	&0.19	&-0.01	&-0.12	&0.26     \\
81809	&5782	&4	&-0.28	&1.3	&-0.12	&0.18	&-0.04	&-0.15	&0.17     \\
107213	&6156	&4.1	&0.07	&1.6	&-0.07	&0.15	&-0.01	&0.02	&       \\
139323	&5204	&4.6	&0.19	&0.7	& 0.31	&0.11	&0.32	&0.00	&0.1      \\
139341	&5242	&4.6	&0.21	&0.9	&-0.06	&0.10	&-0.05	&-0.07&	0.13     \\
144579	&5294	&4.1	&-0.7	&1.3	& 0   	&0.21	&0.11	&-0.25	& 	0.24     \\
159222	&5834	&4.3	&0.06	&1.2	&-0.09	&0.15	&-0.03	&-0.03	&-0.07    \\
159909	&5749	&4.1	&0.06	&1.1	&-0.13	&0.14	&-0.08	&-0.11	&-0.03    \\
215704	&5418	&4.2	&0.07	&1.1	&-0.03	&0.13	&0.00	&-0.12	&-0.03    \\
218209	&5705	&4.5	&-0.43	&1	&-0.08	&0.19	&0.01	&-0.01	&-0.03    \\
221354	&5242	&4.1	&-0.06	&1.2	&-0.12	&0.14	&-0.07	&-0.26&-0.03    \\

\hline 
 nonclassified   &  &  &  &  &  &  &  &  &    \\
\hline
4628	&4905	&4.6	&-0.36	&0.5	& -0.09	&0.16	&-0.02	&-0.04	&         \\
4635	&5103	&4.4	&0.07	&0.8	&  0.03	&0.12	&0.05	&-0.04	&0        \\
10145	&5673	&4.4	&-0.01	&1.1	& -0.11	&0.15	&-0.05	&-0.06	&0.15     \\
12051	&5458	&4.55	&0.24	&0.5	& -0.04	&0.11	&-0.02	&0.1	&-0.07    \\
13974	&5590	&3.8	&-0.49	&1.1	& -0.11	&0.19	&-0.01	&-0.01		&0.01     \\
17660	&4713	&4.75	&0.17	&1.3	&  0.03	&0.10	&0.03	&-0.14		&0.15     \\
20165	&5145	&4.4	&-0.08	&1.1	& -0.02	&0.14	&0.02	&-0.07		&0        \\
24206	&5633	&4.5	&-0.08	&1.1	& -0.04	&0.15	&0.018	&0.03	&0.07     \\
32147	&4945	&4.4	&0.13	&1.1	&  0.27	&0.11	&0.28	&-0.04		&0.06     \\
45067	&6058	&4	&-0.02	&1.2	& -0.1 	&0.16	&-0.03	&0		&-0.04    \\
84035	&4808	&4.8	&0.25	&0.5	&  0.11	&0.09	&0.10	&-0.05	&-0.08    \\
86728	&5725	&4.3	&0.22	&0.9	& -0.07	&0.13	&-0.03	&-0.06	&-0.1     \\
90875	&4788	&4.5	&0.24	&0.5	&  0.26	&0.09	&0.25	&-0.01	&         \\
117176	&5611	&4	&-0.03	&1	& -0.07	&0.15	&-0.01	&-0.01	&0.07     \\
117635	&5230	&4.3	&-0.46	&0.7	& -0.02	&0.18	&0.06	&-0.04	&0.3      \\
154931	&5910	&4	&-0.1	&1.1	& -0.09	&0.16	&-0.02	&0.01		&-0.01    \\
159482	&5620	&4.1	&-0.89	&1	& -0.03	&0.23	&0.10	&-0.01	&0.35     \\
168009	&5826	&4.1	&-0.01	&1.1	& -0.09	&0.15	&-0.03	&-0.06		&0.05     \\
173701	&5423	&4.4	&0.18	&1.1	& -0.01	&0.12	&0.01	&-0.1		&-0.14    \\
182736	&5430	&3.7	&-0.06	&1	&  0.03	&0.14	&0.07	&0.05		&         \\
184499	&5750	&4	&-0.64	&1.5	& -0.04	&0.21	&0.07	&-0.1	&0.37     \\
184768	&5713	&4.2	&-0.07	&1.1	& -0.09	&0.15	&-0.03	&-0.09	&0.11     \\
186104	&5753	&4.2	&0.05	&1.1	& -0.07	&0.15	&-0.02	&-0.05		&0.09     \\
215065	&5726	&4	&-0.43	&1.1	& -0.15	&0.19	&-0.05	&-0.16		&         \\
219134	&4900	&4.2	&0.05	&0.8	&  0.05	&0.11	&0.06	&-0.03	& 	-0.11    \\
219396	&5733	&4	&-0.1	&1.2	& -0.1 	&0.16	&-0.03	&-0.09		&         \\
224930	&5300	&4.1	&-0.91	&0.7	& -0.04	&0.24	&0.10	&-0.09	&         \\
\hline
\end{longtable} 

\label{lastpage}

\bsp

\end{document}